\documentclass[letterpaper, 10 pt, conference]{ieeeconf}  
\IEEEoverridecommandlockouts                        
\makeatletter
\newcommand{\linebreakand}{%
  \end{@IEEEauthorhalign}
  \hfill\mbox{}\par
  \mbox{}\hfill\begin{@IEEEauthorhalign}
}
\makeatother

\usepackage{graphicx}
\usepackage{comment}
\usepackage[utf8]{inputenc}
\usepackage{array}
\usepackage[usenames,dvipsnames]{color}
\usepackage{multirow}

\usepackage [ T1 ] { fontenc }
\usepackage [ linguistics ] {forest}

\newcolumntype{L}{>{\centering\arraybackslash}m{10cm}}

\usepackage[T1]{fontenc}
\usepackage{tikz}
\usepackage{adjustbox}
\usetikzlibrary{arrows,calc,positioning}
\usetikzlibrary{backgrounds,fit}

\usetikzlibrary{matrix,fit}

\tikzstyle{intt}=[draw,text centered,minimum size=6em,text width=5.25cm,text height=0.34cm]
\tikzstyle{intl}=[draw,text centered,minimum size=2em,text width=2.75cm,text height=0.34cm]
\tikzstyle{int}=[draw,minimum size=2.5em,text centered,text width=3.5cm]
\tikzstyle{intg}=[draw,minimum size=3em,text centered,text width=6.cm]
\tikzstyle{sum}=[draw,shape=circle,inner sep=2pt,text centered,node distance=3.5cm]
\tikzstyle{summ}=[drawshape=circle,inner sep=4pt,text centered,node distance=3.cm]

\title{\LARGE \bf
The Threat of Adversarial Attacks Against Machine Learning in Network Security: A Survey
}

\author{ \parbox{5 in}{\centering Olakunle Ibitoye, Rana Abou-Khamis, Mohamed el Shehaby, Ashraf Matrawy and  M. Omair Shafiq 
         \thanks{}\\
        Carleton University, Ottawa, Canada\\
         \tt\small Emails: Kunle.Ibitoye@carleton.ca, Rana.Aboukhamis@carleton.ca, 
         MohamedelShehaby@cmail.carleton.ca,
         Ashraf.Matrawy@carleton.ca, Omair.Shafiq@carleton.ca
}
}

\begin{document}

\maketitle



\begin{abstract}

Machine learning models have made many decision support systems to be faster, more accurate and more efficient. However, applications of machine learning in network security face more disproportionate threat of active adversarial attacks compared to other domains. This is because machine learning applications in network security such as malware detection, intrusion detection, and spam filtering are by themselves adversarial in nature. In what could be considered an arm’s race between attackers and defenders, adversaries constantly probe machine learning systems with inputs which are explicitly designed to bypass the system and induce a wrong prediction. In this survey, we first provide a taxonomy of machine learning techniques, tasks, and depth. We then introduce a classification of machine learning in network security applications. Next, we examine various adversarial attacks against machine learning in network security and introduce two classification approaches for adversarial attacks in network security. First, we classify adversarial attacks in network security based on a taxonomy of network security applications. Secondly, we categorize adversarial attacks in network security into a problem space vs. feature space dimensional classification model. We then analyze the various defenses against adversarial attacks on machine learning-based network security applications. We conclude by introducing an adversarial risk grid map and evaluate several existing adversarial attacks against machine learning in network security using the risk grid map. We also identify where each attack classification resides within the adversarial risk grid map.

\end{abstract}


\section{Introduction}\label{intro}

There has been an ever-increasing application of machine learning and deep learning techniques in network security. One key advantage of machine learning is that it makes optimal decisions more feasible.

It, however, introduces a new challenge since security and robustness of these models is usually not a huge consideration for machine learning algorithm designers who are more focused on designing effective and efficient models. This creates room for various forms of attack models against machine learning-based network security applications.

Researchers \cite{goodfellow6572explaining}\cite{kos2018adversarial}\cite{moosavi2016deepfool}\cite{papernot2016limitations} have shown that the presence of adversarial samples can easily fool machine learning systems. Adversarial samples are specially crafted inputs that cause a machine learning model to classify an input wrongly. Machine learning systems typically take in input data in two distinct phases. The training data which is fed into the learning algorithm during the training phase, and the new or test data which is fed into the learned model during the prediction phase. If the attacker can manipulate the input data in either phase, it is possible to induce a wrong prediction from the machine learning model.

In this survey, we provide a brief introduction to machine learning using a three-dimensional classification method. We classify the various machine learning approaches based on the learning tasks, learning techniques and learning depth. We further organize the various applications of machine learning in network security based on a taxonomy of security tasks.  Contrary to the survey by Corona et al. \cite{corona2013adversarial}, our work focuses on adversarial attacks that are strictly machine learning based. 
Next, we classify the various adversarial attacks based on the applications in network security. We identify five main categories of machine learning applications in network security for our classification method. Finally, we classify adversarial attacks against machine learning based on a taxonomy of network security applications.

Our \textbf{contribution} is threefold. First, we introduce a new method for classifying adversarial attacks in network security based on a taxonomy of network security applications. We also introduce the concept of problem space and feature space dimensional classification of adversarial attacks in network security.

Secondly, we introduce the concept of adversarial risk in computer and network security. We provide a new risk mapping for evaluating the risk of adversarial attacks in network security based on the discriminative or directive autonomy of the machine learning tasks and techniques respectively. 


Lastly, we evaluate several adversarial attacks against machine learning in network security applications as proposed by various researchers and classify the attacks based on an adversarial threat attack taxonomy shown in Table \ref{Table:Adversarial_Attack_Taxonomy}. 


As we outline in Section \ref{related}, prior adversarial attacks surveys \cite{madry2017towards}\cite{papernot2017practical}\cite{barreno2006can} mainly covered them in the computer vision domain. Nevertheless, some surveys tackled adversarial attacks on cybersecurity \cite{rosenberg2021adversarial}\cite{duddu2018survey}\cite{zhang2019deep}\cite{biggio2018wild}, but to the best of our knowledge, there is currently no prior work that has reviewed adversarial attacks in network security based on a classification of network security applications. No prior work has also reviewed the concept of problem space vs. feature space dimensional classification of adversarial attacks in network security.  Also, this is the first work to propose an adversarial machine learning risk grid map in the field of network security based on the directive or discriminative autonomy of the machine learning algorithms.


\begin{figure}[h]
\includegraphics[width=\linewidth,keepaspectratio=true]{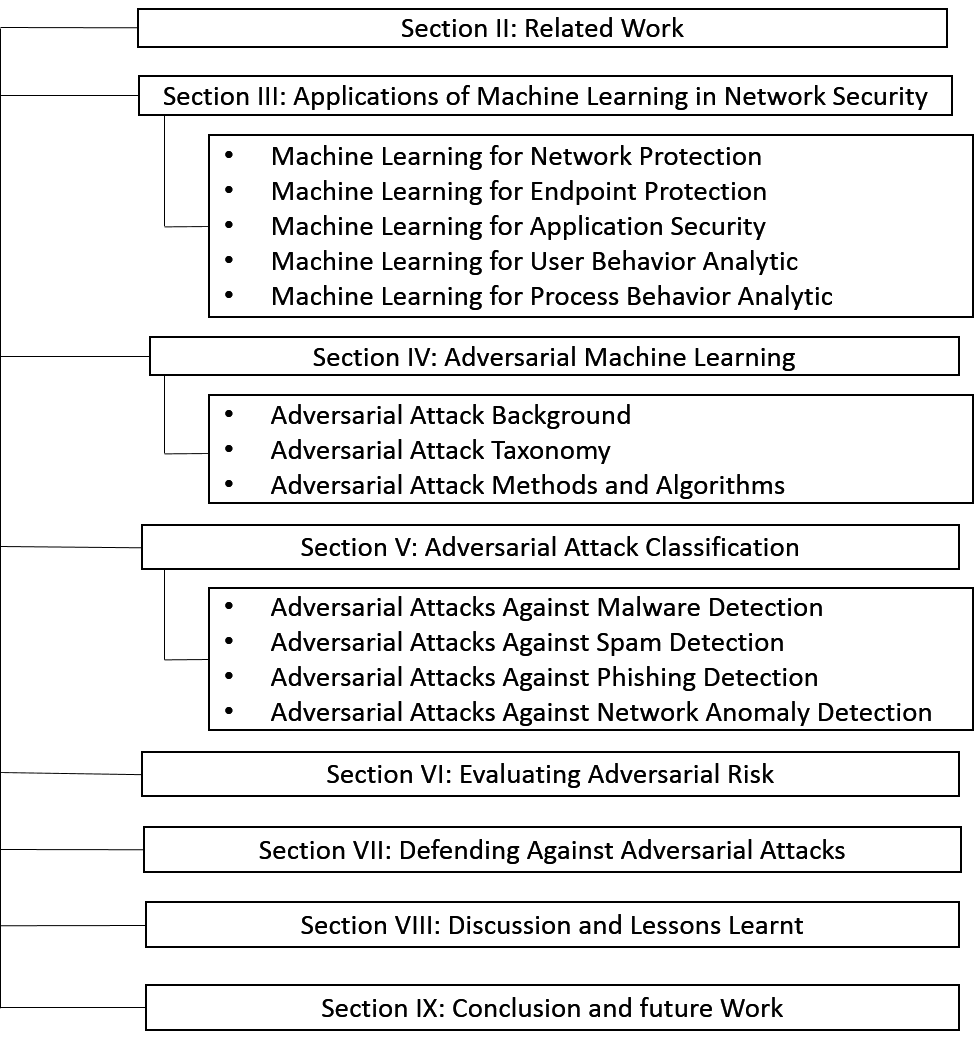}
  \caption{Structure of the Paper}
  \label{fig:PaperStructure}
  \centering
\end{figure}

As illustrated in Figure \ref{fig:PaperStructure}, We structure the remainder of the paper as follows. In Section II, we survey some related work. In section III, we discuss 
some applications of machine learning in network security. In section IV, we begin with a brief background about adversarial machine learning followed by  a description of our adversarial attack taxonomy. We also review different adversarial attack methods and algorithms. In section V, we introduce a classification method for adversarial attacks in network security based on the network security CIA goals of confidentiality, integrity and availability. In section VI, we discuss and evaluate adversarial risk in machine learning. In section VII, we review various approaches for defending against adversarial attacks. In section VIII, we provide some discussion and lessons learnt. Finally, in section IX, we add a conclusion for our survey with guidance for future work.

\section{Related Work}\label{related}

Adversarial attacks have been widely studied in the field of computer vision \cite{madry2017towards}\cite{papernot2017practical}\cite{barreno2006can} with several attack methods and techniques developed mostly for image recognition tasks. Researchers have discussed the public safety concern of adversarial attacks such as in self-driving cars which could be fooled into mis-classifying a stop sign resulting in a potentially fatal outcome \cite{amodei2016concrete}. In network security, the consequences of adversarial attacks are equally significant \cite{vorobeychik2018adversarial} especially in areas such as intrusion detection \cite{wang2018deep} and malware detection \cite{liu2018adversarial} where there have been rapid progress in the adoption of machine learning for such tasks. Even though adversarial machine learning has recently been widely researched in network security, to the best of our knowledge, there is currently no publication that has surveyed the vast number of growing research work on adversarial machine learning in this field. Some existing survey papers we reviewed include Akhtar et al. \cite{akhtar2018threat} which reviewed adversarial attacks against deep learning in computer vision. Qui et al \cite{qiu2019review} provided a generalized survey on adversarial attacks in artificial intelligence, with a brief discussion on cloud security, malware detection and intrusion detection. Liu et al. \cite{liu2018survey} reviewed security threats and corresponding defensive techniques of machine learning focusing on the threats in the learning algorithms. Rosenberg et al. \cite{rosenberg2021adversarial} provided a general review on adversarial attacks on cyber security domains like; Intrusion detection systems, URL Detection systems, Biometric Systems, CPSs (Cyber-Physical Systems), and Industrial Control Systems. Unlike their work, our review only concentrates on network security and uses different approaches to classify adversarial attacks and defenses. Duddu el al. in \cite{duddu2018survey} discussed various research work on adversarial machine learning in cyberwarfare, with some mention of adversarial attacks against malware classifiers. Zhang et al. \cite{zhang2019deep} discussed adversarial attacks as a limitation of deep learning in mobile and wireless networking but did not consider deep learning in the context of network security applications. Buczak et al. \cite{buczak2015survey} in their survey on machine learning-based cybersecurity intrusion detection focused on complexity and challenges of machine learning in cybersecurity but did not review adversarial attacks in their study. Biggio and Roli \cite{biggio2018wild} provided an historical timeline of adversarial machine learning in the context of computer vision and cybersecurity but their work did not provide a detailed review in the context of network security. Gardiner et al. \cite{gardiner2016security} in their survey on the security of machine learning in malware detection, focused on reviewing the Call and Control (C \& C) detection techniques. They also identified the weaknesses and explained the limitations of secure machine learning algorithms in malware detection systems. Domain specific surveys on adversarial machine learning has also been published including Hao et al. \cite{hao2020adversarial} in which various adversarial attacks and defenses in images, graphs and texts were reviewed. In the field of natural language processing, zhang et.al \cite{zhang2020adversarial} reviewed various publications in which deep adversarial attacks and defenses were proposed. Sun et al. \cite{sun2018survey} published a survey on adversarial machine learning in graph data. Akhtar et al. \cite{akhtar2018threat} computer vision, Duddu et al. \cite{duddu2018survey} cyber warfare.\newline

\textbf{Research Gap}
With growing interest in the use of machine learning for network security applications, the significance of adversarial attacks against such machine learning-based application have become more prevalent. With continued increase in the amount of work in this field, there have been recent attempts to review these publications into a survey work. In the field of network security, We identified nine survey papers which attempt to discuss adversarial machine learning from the context of network security. None of these previous survey papers have however explored the vast amount of research work currently ongoing on the topic of adversarial machine learning in network security in a manner that categorizes them based on security applications, problem and feature space dimensional classification and adversarial risk grid map.

Our survey more importantly seeks to distinguish between adversarial attacks in general, and adversarial machine learning in context. We note that an adversary may seek to compromise network security applications in various ways and this may not be related to adversarial machine learning. For example in \cite{corona2013adversarial} where adversarial attacks in Intrusion detection systems was reviewed. In our context, adversarial machine learning specifically addresses the optimization problem in which a machine learning based network security solution is being attacked. Many network security solutions are strictly rules based or hard programming dependent and do not implement machine learning techniques. Our survey work does not refer to such adversarial attacks, since they do not capture the real context of adversarial machine learning in principle.

\section{Applications of Machine Learning in Network Security}

\begin{figure*}[h!]
\centering
\includegraphics[width=0.95\linewidth,keepaspectratio=true]{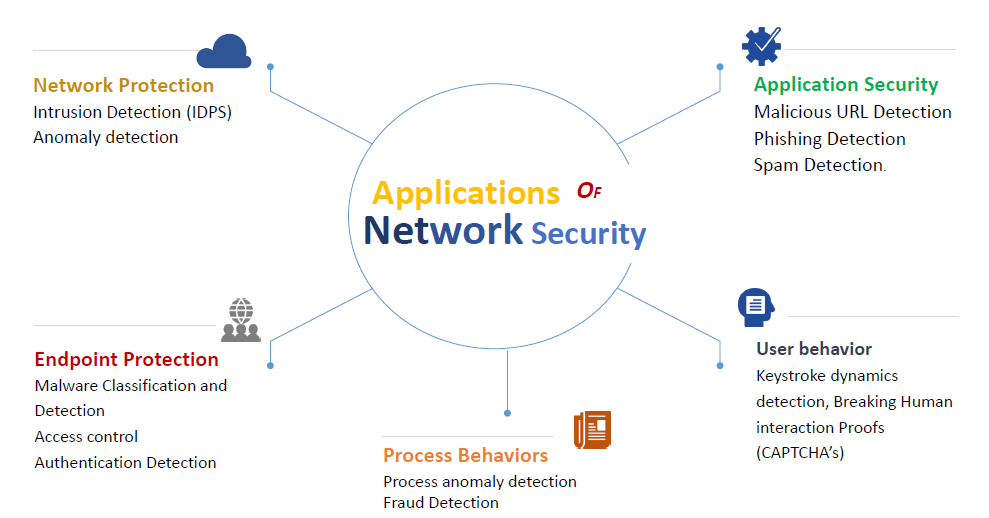}
  \caption{Machine Learning Applications in Network Security}
  \label{fig:net_applications}
  \centering
\end{figure*}

Today's network as well as next generation network architectures have become quite complex, and new innovations of network security solutions are required to protect against the growing landscape of cyber threats. Machine learning techniques have been increasingly used to carry out a wide range of tasks in network security \cite{ford2014applications} incorporating several layers of defenses both within the network and at the edge of the network.  In this section, we review and highlight some applications of machine learning in network security by classifying them into five categories as illustrated in Figure \ref{fig:net_applications}.

\subsection{Machine Learning for Network Protection} 
Intrusion Detection Systems (IDS) are essential solutions for monitoring events dynamically in a computer network or system. Essentially there are two types of IDS (signature based and anomaly based) \cite{singh2013survey}. Signature based IDS detects attacks based on the repository of attacks signatures with no false alarm \cite{almseidin2017evaluation}. However, zero-day attacks can easily bypass signature-based IDS. Anomaly IDS \cite{almseidin2017evaluation} uses machine learning and can detect a new type of attacks and anomalies. A typical disadvantage of anomaly IDS is the tendency to generate a significant number of false positive alarms.

\begin{itemize}

\item{Hybrid Approach for Alarm Verification} 
Sima et al. \cite{sima2018hybrid} designed and built Hybrid Alarm Verification System that requires processing a significant number of real-time alarms, high accuracy in classifying false alarms, perform historical data analysis. The proposed system consists of three components:  Machine Learning, Stream processing and Batch processing (Alarm History). Machine learning model trained offline and used for verification service that can immediately classify true or false alarms. They used different machine learning algorithms in the experiments to show the effectiveness of their system where the accuracy achieves more than 90\% in a stream of 30K alarms per second \cite{sima2018hybrid}.

\item{Learning Intrusion Detection}
Laskov et al. \cite{laskov2005learning} worked in developing a framework to compare the supervised learning (classiﬁcation) and unsupervised learning (clustering) techniques for detecting intrusions and malicious.
They used different methods in supervised learning to evaluate the work include k-Nearest Neighbor (kNN), decision trees, Support Vector Machines (SVM) and Multi-Layer Perception (MLP). Also, k-means clustering was utilized, with single linkage clustering as unsupervised algorithms. The evaluation was ran under two scenarios to evaluate how much the IDS could generalize its knowledge to new malicious activities. The supervised algorithms showed better classification with the known attacks. The best result among the supervised algorithm was the decision tree algorithm whiched achieved  95\% true positive and 1\% false positive rate, followed by MLP, SVM and then KNN. If there were new attacks not previously seen in the training data, the accuracy decreases significantly. However, the unsupervised algorithms performed better for unseen attacks and did not show signiﬁcant diﬀerence in accuracy for seen and unseen attacks \cite{laskov2005learning}.
 
\end{itemize}

\subsection{Machine Learning for Endpoint Protection}

Malware detection is a significant part of endpoint security including workstations, servers, cloud instances, and mobile devices. Malware detection is used to detect and identify malicious activities caused by malware. With the increase in the variety of malware activities, the need for automatic detection and classifier amplifies as well. The signature-based malware detection system is commonly used for existing malware that has a signature but it not suitable for unknown malware or zero-day malware. Machine learning can cope with this increase and discover underlying patterns in large-scale datasets \cite{kolosnjaji2016deep}.
\begin{itemize}

\item{Automatic Analysis of Malware Behavior}
Rieck et al. \cite{rieck2011automatic} successfully proposed a framework for analyzing malware behavior automatically using various machine learning techniques. The framework allows clustering similar malware behaviors into classes and assigns new malware to these discovered classes. They designed an incremental approach for the behavior analysis that can process various malware behaviors and reduce the run-time defense against malware development comparing to other analysis methods and provide accurate discovery of novel malware. To implement this automatic framework, they collected a large number of malware samples and monitored their behaviors using a sandbox environment and learn those behaviors using Clustering and Classification algorithms \cite{rieck2011automatic}.

\item{Automated Multi-level Malware Detection System}
In \cite{kumara2018automated}, authors proposed Advanced Virtual Machine Monitor-based guest-assisted Automated Multilevel Malware Detection System (AMMDS) that affect both Virtual Machine Introspection (VMI) and Memory Forensic Analysis (MFA) techniques to mitigate in real time symptoms of stealthily hidden processes on guest OS \cite{kumara2018automated}. They use different machine learning techniques such as Logistic Regression, Random Forest, Naive Bayes, Random Tree, Sequential Minimal Optimization (SMO), and J48 to evaluate the AMMDS and the results achieve 100\%.

\item{Classiﬁcation of Malware System Call Sequences}
Kolosnjaji et al. \cite{kolosnjaji2016deep} focused on the utilization of neural networks by stacking layers according to deep learning to improve the classiﬁcation of newly retrieved malware samples into a predeﬁned set of malware classes. They constructed Convolutional Neural Network (CNN) and Recurrent Neural Network (RNN) layers for modeling System Call Sequences. The sequences used by the CNN layers was based on a set of n-grams. The presence of the n-grams and  their relation were counted in a behavioral trace. The RNN on the other hand  used sequential information to train the model. A dependence between the system call appearance and the system call sequence was however maintained. If this model was trained properly, it usually provided better accuracy on subsequent data and most often captured more training set information. This deep learning technique for capturing the relation between the n-grams in the system call sequences was deemed to be relatively efficient as it achieved 90\% average accuracy, precision and recall for most of the malware families \cite{kolosnjaji2016deep}. 

\item{A Hybrid Malicious Code Detection Method}
Li et al. \cite{li2015hybrid} proposed a hybrid malicious code detection scheme based on AutoEncoder and Deep Belief Networks (DBN). They used the AutoEncoder to reduce the dimensionality of data by extracting the main features. Then they used the DBN that composed multilayer Restricted Boltzmann Machines (RBM) and a layer of BP neural network to detect malicious code. The BP neural network has an input vector from the last layer of RBM based on unsupervised learning and then use supervised learning in the BP neural network. 
They achieved the Optimal hybrid model. The experiment results that are verified by KDDCUP'99 dataset show higher accuracy compared to a single DBN and reduce the time complexity \cite{li2015hybrid}. 

\end{itemize}

\subsection{Machine Learning for Application Security}
Various machine learning tasks used for application security including malicious web attack detection, phishing detection and spam detection.

\begin{itemize}

\item{Detection of Phishing Attacks}
Basnet et al. \cite{basnet2008detection} studied and compared the effectiveness of using different machine learning algorithms for classification of phishing emails using many novel input features that helps in detecting phishing attacks. The training dataset is labeled with phishing or legitimate email. They used unsupervised learning to extract features without prior training directly and provides fast and reliable knowledge from the dataset. They used 4000 emails in total, A total of 2000 emails used for testing. They used  Support Vector Machines (SVM), Leave One Model Out, Biased SVM, Neural Networks, Self Organizing Maps (SOMs) and K-Means on the dataset.  Consistently,  Support Vector Machine achieved the best results. The Biased Support Vector Machine (BSVM) and NN have an accuracy of 97.99\% \cite{basnet2008detection}.

\item{Adaptively Detecting Malicious Queries in Web Attacks}
Don et al. \cite{dong2018adaptive} proposed a new system called AMODS and learning strategy called SVM HYBRID for detecting web attacks. AMODS is an adaptive system that aims to periodically update the detection model to detect the latest web attacks. The SVM HYBRID is an adaptive learning strategy which was implemented primarily for reducing manual work. The detection model was trained using dataset which was obtained from an academic institute’s web server logs. The proposed detection model outperformed existing web attack detection methods with an FP rate of 0.09\% and 94.79\% F-value. The SVM Hybrid system obtained a total number of malicious queries equal to 2.78 times by the popular SVM method. Also, the Web Application Firewall (WAF) can use malicious queries to update the signature library. The significant queries were used for updating the detection model which consisted of a meta-classifier as well as other three base classifiers \cite{dong2018adaptive}. 

\item{URLNet -Learning a URL Representation with Deep Learning for Malicious URL Detection}
Le et al. \cite{le2018urlnet} proposed an end-to-end deep learning framework which did not require sophisticated feature. URLNet was introduced to address several limitations which was found with the other model approaches. This framework learns from the URL directly how to perform a nonlinear URL embedding which then enabled it to successfully detect various Malicious URLs. Convolutional Neural Networks (CNN) were applied to both the characters and words of each URL to discover the URL embedding method. They also proposed advanced word-embedding techniques to deal with uncommon words, which was a limitation being experienced by other malicious URL detection systems. The framework then learns from unknown works at testing phase \cite{le2018urlnet}.

\end{itemize}

\subsection{Machine Learning for User Behavior Analytic}
User behavior analytics is a cybersecurity process which involves analyzing patterns in human behaviors and detecting anomalies that give an indication of fraudulent activities or insider threats. Machine learning algorithms are used to detect such anomalies in user actions such as unusual login tries and to infer useful knowledge from those patterns.

\begin{itemize}
\item{Authentication with Keystroke Dynamics}
Revett et al. \cite{revett2007machine} proposed a system using Probabilistic Neural Network (PNN) for keystroke dynamics that captures the typing style of a user. A system comprising of 50 user login credential keystrokes was evaluated. The authors \cite{revett2007machine} used eight attributes to monitor the enrollment and authentication attempts. An accuracy of 90\% was obtained in classifying legitimate users from imposters. A comparison of the training time between the PNN system and a Multi-Layer Perception Neural Network (MLPNN) showed that the PNN was four times faster. 

\item{Text-based CAPTCHA Strengths and Weaknesses}
Bursztein et al. \cite{bursztein2011text} in a study showed that several well known websites still implemented technologies that have been proven to be vulnerable to cyber attacks. In the study, an automated Decaptcha tool was tested on numerous websites including well known names such as eBay, Google and Wikipedia. It was observed that 13 out of 15 widely used web technologies were vulnerable to their automated attack. They had a significant success rate for most of the websites. Only Google and Recaptacha were able to resist to the automated attack. Their study revealed the need for more robust CAPTCHA designs in most of the widely used schemes.
Authors recommended that the schemes should not rely on segmentation alone because it did not provide sufficient defense against automated attacks.

\begin{figure*}[h!]
    \centering
    \includegraphics[width=0.95\linewidth,keepaspectratio=true]{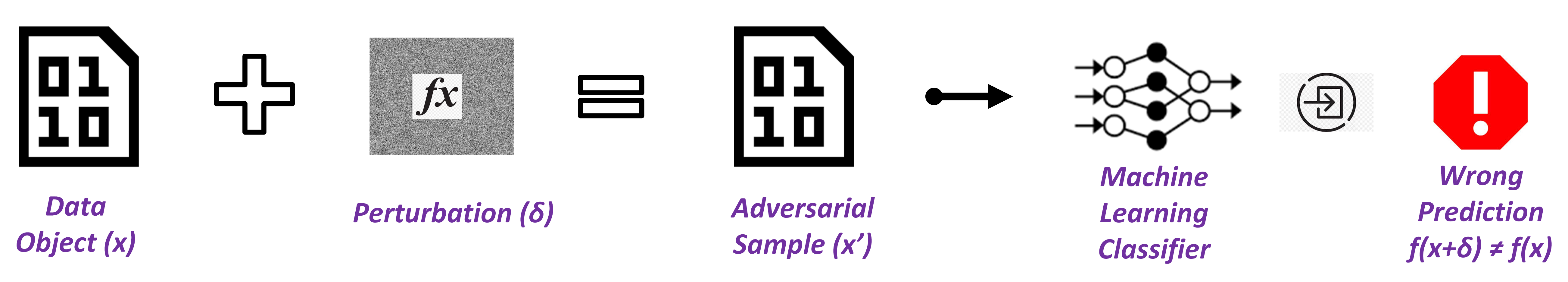}
    \caption{Adversarial Machine Learning}
    \label{fig:rana_snip}
\end{figure*}

\item{ Social Network Spam Detection}
K. Lee et al. \cite{lee2010uncovering} proposed social network spam detection that gathers legitimate and spam profiles and feeds them to Support Vector Machine (SVM) model. The authors selected two social networks: Twitter and MySpace to evaluate the proposed machine learning system. They collected data over months and feed them to the SVM classifier. The dataset contains 388 legitimate profiles and 627 spam profiles collected from MySpace, and 104 legitimate profiles and 168 profiles between promoters and spammers collected from Twitter. The system achieved a low false positive rate and high precision up to 70\% for MySpace and 82\% for Twitter.
\end{itemize}

\subsection{Machine Learning for Process Behavior Analytic}
Machine learning applications usually necessitate the need to learn and have some domain knowledge about business process behaviors in order to detect anomalous behaviors. Machine learning could be used for determining fraudulent transactions within banking systems. Also it has been successfully used for identifying outliers, classifying types of fraud and for clustering various business processes.
\begin{itemize}

\item{Anomaly detection in Industrial Control Systems}
Kravch et al.\cite{kravchik2018anomaly} performed a successful study on SecureWater Treatment Testeb (SWat) using Deep Convolutional Neural Networks CNN to detect most of attacks on Industrial Control System (ICS) with a low false positive. The anomaly detection method was based on the statistical deviation measurement of the predicted value. They performed the study using 36 different attacks from SWat. The authors in \cite{kravchik2018anomaly} proofed that using 1D convolutional networks in anomaly detection in ICS outperformed the recurrent networks.

\item{Detecting Credit Card Fraud}
Traditionally, the Fraud Detection System uses old transactions data to predict a new transaction. Fraud Detection System (FDS) should encounter various potential challenges and difficulties to achieve high accuracy and performance \cite{kou2004survey}. The traditional detection method does not solve all problems and challenges including imbalanced data where there is a small chance of transactions are fraudulent. Wrong classification and overlapping data and Fraud detection cost are other major challenges \cite{kou2004survey}. 
Chen et al. \cite{csahin2011detecting} proposed an approach to solving the listed challenges and problems for Credit Card fraud. They introduced a system to prevent fraud from the initial use of credit cards by collecting user data from online questionnaire based on consumer behavior surveys. They used various classifiers models: decision tree (C5.0, CandRT, CHAID) and SVM ( linear and radial basis, Kernels of polynomial, sigmoid). They use three datasets to develop questionnaire-responded transaction (QRT) model to predict new transaction.

\item{Deep Learning Techniques for Side-Channel Analysis}
Prouff et al. \cite{prouff2018study} defined Side-Channel Analysis as a type of  attack that attempts to leak information from a system by exploiting some parameters from the physical environment \cite{prouff2018study}. This attack was utilizing the running-time of some cryptographic computation, especially in the block ciphers. 
The capability of a system to resist side-channel attacks (SCA) requires an evaluation strategy that focuses on deducing the relationship between the device behavior and the sensitivity of the information that is common in classical cryptography. 
The authors in \cite{prouff2018study} focused on proposing an extensive study of using deep learning algorithms in the Side-Channel Analysis. Also, they focused on the hyper-parameters selection to help in designing new deep learning classifier and models. They confirmed that the Convolutional Neural Networks (CNN) models are better in detecting SCA. Their proposal system outperformed the other tested models on highly desynchronized traces and had the best performance as well on small desynchronized trace \cite{prouff2018study}.
 
\end{itemize}

\section{Adversarial Machine Learning}

Adversarial attacks have been studied for more than a decade now \cite{biggio2018wild}. However, the first notable discovery in adversarial attacks for computer vision was by Szegedy et al. \cite{szegedy2013intriguing} who reported that a small perturbation in the form of a carefully crafted input could confuse a deep neural network to misclassify an image object. Other researchers have demonstrated the use of adversarial attacks beyond image classification \cite{chen2017secmd}\cite{chen2018droideye}\cite{chen2017adversarial}\cite{kolosnjaji2018adversarial}.

In adversarial machine learning, an adversary seeks to confuse a machine learning model into making a wrong decision. The adversary achieves this by modifying the input data that is fed to the machine learning model either during the training phase (poisoning attack) \cite{munoz2017towards} or during the inference phase (evasion attack) \cite{biggio2013evasion}. 

The reason behind adversarial examples has been linked to the fact that most machine learning models remain overtly attached to the superficial statistics of the input data \cite{jo2017measuring} \cite{hendrycks2018benchmarking} .  This attachment to the input data makes the machine learning highly sensitive to distribution shift, resulting in a disparity between semantic changes and a decision change \cite{goodfellow6572explaining}.

We consider the security model for use of machine learning in network security as a combination of four components namely the attack surface, threat model, adversarial framework and adversarial risk. An alternative adversarial model was proposed in \cite{huang2011adversarial} which modeled the adversary using a threefold approach based on knowledge, goals and capability. The attack surface identifies the various attack vectors along a typical machine learning data processing pipeline in network security related applications. The threat model provides a system abstraction for profiling the adversary's capabilities and the potential threats that are associated. The adversarial framework details our approach for classifying the various attacks and defenses within each network security domain and lastly the adversarial risk provides an evaluation of the likelihood and severity of adversarial attacks within a network security system.

A major component of an adversarial attack is the adversarial sample. As illustrated in Figure \ref{fig:rana_snip}, an adversarial sample consists of an input to a machine learning model which has been perturbed. For a particular dataset with features x and label y, a corresponding adversarial sample is a specific data point x' which causes a classifier c to predict a different label on x' other than y, but x' is almost indistinguishable from x. The adversarial samples are created using one of many optimization methods known as adversarial attack methods. Crafting adversarial samples involves solving an optimization problem to determine the minimum perturbation which maximizes the loss for the neural network

Considering an input x, and a classifier f, the optimization goal for the adversary is to compute such perturbation with a s mall norm, measured w.r.t some distance metric, that would modify the output of the classifier such that \[f(x + \delta) \neq f(x)\] where \(\delta\) is the perturbation. If \(\delta\) is applied to all of the input data (all of the image's pixels, for example), it is considered a \textbf{dense adversarial attack}. However, if just partial positions are perturbed, it is called a \textbf{sparse adversarial attack} \cite{fan2020sparse}.

Adversarial machine learning in network security is typically an arms race between two agents. The first agent is an adversary whose objective is to intrude a network with a malicious payload. The other agent is one whose role is to protect the network from the consequences of the malicious payload.

We start with a view of the different type of data that traverses a network during any given time.

\subsection{Adversarial Attack Taxonomy}\label{threat_model}

\begin{table*}[t]
\small
\caption{Adversarial Attack Taxonomy}
\label{Table:Adversarial_Attack_Taxonomy}
\begin{center}
  \centering
\begin{tabular}{|c|c|c|} 
 \hline
  & Types & References \\  
 
 \hline
  \multirow{3}{*}{Knowledge}  & Black Box & \cite{papernot2017practical} \cite{hu2018black} \\ 
 \cline{2-3}
           & White Box& \cite{zhang2019defending} \cite{goodfellow6572explaining} \\ 
 \cline{2-3}
             & Gray Box & \cite{carlini2019evaluating} \\ 
 \hline
 \multirow{2}{*}{Space} & Feature Space & \cite{wang2018deep}  \\ 
 \cline{2-3}
   & Problem Space & \cite{pierazzi2020intriguing}  \\ 
 \hline
 \multirow{3}{*}{Strategy}  & Evasion & \cite{carlini2019evaluating} \cite{pattanaik2018robust} \\ 
 \cline{2-3}
  & Poisoning  & \cite{tabassi2019taxonomy} \cite{shafahi2018poison} \cite{munoz2017towards}\\
 \cline{2-3}
  &Oracle & \cite{tramer2016stealing} \cite{tabassi2019taxonomy} \\ 
 \hline
 
  \multirow{3}{*}{Goal} & Availability &\cite{papernot2018sok} \cite{kim2020over}\\ 
 
 \cline{2-3}
    & Integrity & \cite{papernot2018sok} \cite{usama2019generative} \\ 
 
 \cline{2-3}
   & Confidentiality & \cite{papernot2018sok} \cite{juuti2019prada} \\ 
 
 \hline
 \multirow{2}{*}{Target} & Physical Domain  & \cite{albaseer2020performance} \cite{sadeghi2019physical}  \\ 
 \cline{2-3}
   & ML Model & \cite{goodfellow6572explaining}  \\ 
 \hline

\end{tabular}
  \end{center}
\end{table*}


We examine the Adversarial Attack Taxonomy in Table \ref{Table:Adversarial_Attack_Taxonomy} to consider the goals and capabilities of any adversary for a machine learning system. We base our threat framework from the original model in \cite{barreno2006can} \cite{huang2011adversarial}  and adapt it within the context of adversarial attacks in network security domain. Within this context, adversarial attack threats in network security may be considered based on the attacker's knowledge, attack space, attacker's strategy, attacker's goal and attack target. As mentioned in section \ref{intro}, to the best of our knowledge, this is the first review to add the idea of the space dimension in the classification of adversarial attacks in network security. \newline\newline

\subsubsection{Knowledge}
The knowledge component of the adversarial threat model describes the extent to which the adversary knows about the machine system as a whole. This could be classified as \textbf{White-box}, \textbf{Gray-box} or \textbf{Black-box} attacks.

\begin{itemize}
    \item In white-box attacks, it is assumed that the attacker has complete knowledge of the training data, the learning algorithm, the learned model as well as the parameters which were used while training model. A white-box attack represents an adversary who has the exact information that is held by the owner or creator of the machine learning system which is being under attack. In the majority of real world adversarial attack settings, this is usually not feasible. 
    
    \item A Gray-box attacks assumes a more realistic approach, and considers that there could be varying degrees information accessible to the adversary \cite{carlini2019evaluating}. For example, an adversary may have partial information about the model queries, or limited access to the training data. For a gray-box attack, the adversary does not have the exact knowledge which the creator of the model possesses, but has sufficient information to attack the machine learning system to cause the machine learning system to fail. 
    
    \item A black-box attack assumes that the adversary is totally unaware of the machine learning system. in this type of attack, the adversary has no knowledge about either the learning algorithm or the learned model. It may be argued that a truly black-box attack is impossible. this is because it is assumed that the adversary must at least have some specific information, for example the location of the model before it can attack the model. The severity of blackbox attacks poses a greater threat in practice. The model for real-world systems may be more restrictive than a theoretical black-box model where the adversary can understand the full output of the neural network on inputs that have been chosen arbitrarily. In \cite{ilyas2018black}, an analysis of three threat models were proposed. These models, defined as, the query-limited setting, the partial information setting, and the label-only setting, provide a more accurate characterization of real-world classifiers. As such, a representation of black box adversarial attacks was proposed, such that, it would be possible to fool classifiers under these more restrictive threat models, whereas, it might have been impractical or ineffective.\newline\newline 
  
\end{itemize}

\subsubsection{Space}
 In the field of adversarial machine learning, the input space can be defined as a dimensional representation of all the possible configurations of the objects in determination context. We categorize this as \textbf{Feature Space} and \textbf{Problem Space}.
 
\begin{itemize}

    \item Feature space modeling of an adversarial sample is a method in which an optimization algorithm is used to find the ideal value out of a finite number of arbitrary changes made to the features. In a feature space adversarial attack, the attacker's objective is to remain benign without generating a new instance. Conversely, a feature space is defined as the n dimensional space in which all variables in the input dataset are represented. We take as an example an intrusion detection dataset with 70 variables, this represents a 70-dimensional feature space. A feature space adversarial attack in the context above will seek to alter the feature space by making changes within the 70-dimensional feature space. A feature space attack modifies the features in the instance directly. Using an example of malware adversarial attacks, a feature space adversarial malware attack will only modify the feature vectors but no new malware is created.   
    
    \item The problem space refers to an input space in which the objects e.g. image, file, etc. resides. A problem space adversarial malware attack will modify the actual instance from the source to produce a new instance of the malware. Typically, a problem space adversarial attack tends to generate new objects in domains such as malware detection whereby there is no clear inverse mapping to the feature space \cite{pierazzi2020intriguing}.  A typical difference between a problem space adversarial attack, and a feature space adversarial attack is that a feature space attack does not generate a new sample but only creates a new feature vector. A problem space adversarial attack modifies the actual instance itself to create an entirely new object.

\end{itemize}

\subsubsection{Strategy}\label{strategy}
Attacker's strategy implies the phases of operation in which the adversary launches the attack. Three main strategies which an adversary may use in adversarial attacks are \textbf{Evasion}, \textbf{Poisoning} and \textbf{Oracle}.
\begin{itemize}
    \item Evasion attacks, also known as exploratory attack or attack at decision time, during the testing or inference phase. The attacker aims to confuse the decision of the machine learning model after it has been learned as shown in Figure \ref{fig:evasion_attack}. Evasion attacks typically involve an arithmetic computation of an optimization problem. The objective of the optimization problem is to compute a tiny perturbation \(sigma\) which would cause an increase in the loss function. The change in loss function would then be significant enough to result in a wrong prediction by the machine learning model. Evasion attacks are classified as gradient-based attacks or gradient-free attacks. 
    
    Gradient-based attacks are further classified based on the frequency with which the adversarial samples are updated or optimized.  These are \textbf{iterative} or \textbf{One-shot} attacks. Iterative attacks provide tighter control of the perturbation in order to generate more convincing adversarial samples \cite{tabassi2019taxonomy}. This however results in higher computational costs. Alternative to iterative attacks are one-shot attacks which adopt a single-step approach without iterations. One-shot or one-time attacks are attacks in which the adversarial samples are optimized just once. Iterative attacks, however, involve updating the adversarial samples multiple times. By updating the adversarial samples multiple times, the samples are better optimized and perform better compared to one-shot attacks. However, iterative attacks cost more computational time to generate. 
    
    Adversarial attacks against certain machine learning techniques which are computationally intensive such as reinforcement learning  usually demand one-shot attacks as the only feasible approach \cite{pattanaik2018robust}. 
    
    Gradient-free attacks \cite{carlini2019evaluating}, unlike gradient-based attacks do not require knowledge of the model. Gradient-free attacks can generate potent attacks against a machine learning model with knowledge of only the confidence values of the model.
    
    \begin{figure}[h]
\includegraphics[width=\linewidth,keepaspectratio=true]{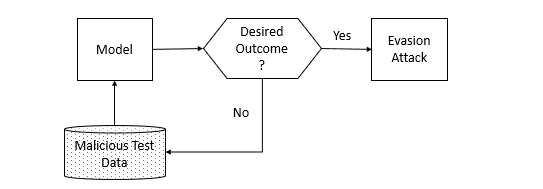}
  \caption{Evasion Attack}
  \label{fig:evasion_attack}
  \centering
\end{figure} 
    
    \item Poisoning attacks, also known as causative attack, involves adversarial corruption of the training data or model logic during the training phase to induce a wrong prediction from the machine learning mode as shown in Figure \ref{fig:poisoning_attack}. Poisoning attacks may be carried out by data injection, data manipulation or logic corruption \cite{tabassi2019taxonomy}. Data injection occurs when the adversary inserts adversarial inputs to alter the data distribution while preserving the original input features and data labels. Data manipulation refers to a situation in which either the input features or data labels of the original training data are modified by the adversary. Logic corruption is an attempt by the adversary to model structure. \newline\newline
    
    \begin{figure}[h]
\includegraphics[width=\linewidth,keepaspectratio=true]{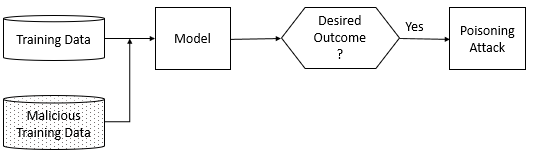}
  \caption{Poisoning Attack}
  \label{fig:poisoning_attack}
  \centering
\end{figure}
    
\end{itemize}

\begin{itemize}
    \item Oracle attacks occur when an adversary leverages the access to the Application Programming Interface of a model, to create a substitute model with malicious intent. The substitute model typically preserves a significant part of the functionality of the original model \cite{tramer2016stealing}. As a result, the substitute model can then be used for other types of attacks such as evasion attacks \cite{tabassi2019taxonomy}. Oracle attacks can be further subdivided into \textbf{Extraction}, \textbf{Inversion} and \textbf{Inference} attacks. The objective of an extraction attack is to deduce model architectural details such as parameters and weights from an observation of the model's output predictions and class probabilities \cite{jagielski2020high}. Inversion attacks occurs when adversary attempts to reconstruct the training  data. An inference attacks allows the adversary to identify specific data points with the distribution of the training dataset \cite{fredrikson2015model}.

\end{itemize}

\subsubsection{Goal}
Traditionally in the field of computer vision, adversarial attacks are regarded in terms of targeted or reliability attacks \cite{akhtar2018threat}. In targeted attacks, the attacker has a specific goal with regard to the model decision. Most commonly, the attacker would aim to induce a definite prediction from the machine learning model. On the other hand, a reliability attack occurs when the attacker only seeks to maximize the prediction error of the machine learning model without necessarily inducing a specific outcome. Yevgeny et al. \cite{vorobeychik2018adversarial} have noted that the distinction between reliability and targeted attacks becomes blurred in attacks on binary classification tasks such as malware binary classification. As such, these conventional paradigms of attacker goal classification is not optimal for consideration in network security. We choose to adopt the CIA triad in this context and find that it is more suitable for adversarial classification  of the adversary goals in network security domain.

\begin{itemize}
    \item Confidentiality attack refers to the goal of the attacker to intercept communication between two parties \(A\) and \(B\), to gain access to private information being exchanged. This happens within the context of adversarial machine learning, whereby machine learning techniques are being used to carry out network security tasks.
    
    \item Integrity attack seeks to cause a misclassification, different from the actual output class which the machine learning model was trained to predict. Integrity attack could result in a targeted misclassification or a reliability attack. A targeted misclassification attempts to make the machine learning model to produce a specific wrong prediction. A reliability attack results in either a confidence reduction or a misclassification to any arbritrary class apart from the correct class.
    
    \item Availability Attack results in a denial of service situation for the machine learning model. as a result, the machine earning model becomes either totally unavailable to the user, or the quality is significantly degraded to the extent that the machine learning system becomes unusable to the end users. 
    
\end{itemize}

\subsubsection{Target}

In our surveyed work, adversarial attacks are targeted against a specific machine learning technique. Several successful attempts have been made towards the transferability of adversarial attacks \cite{papernot2016transferability} \cite{liu2016delving}. However, attacks that have been targeted towards a specific machine learning technique for example unsupervised learning, have not been successfully transfered towards a another technique for example supervised learning. Regarding the physical domain, it includes input sensors, cameras and output actions.

\subsection{Adversarial Attack Methods and Algorithms} 
We recall that adversarial attacks could be deployed either during decision time (evasion attacks) or during training time (poisoning attacks). In each case, the training algorithm (for poisoning attacks) or the learned model (for evasion attacks) is being manipulated with some form of carefully crafted input known as the adversarial samples. A common trend among the attack methods below reveals that the robustness of a machine learning model to a large extent depends on the ability of an attacker to find an adversarial sample that is as close as possible to the original input. In this section, we evaluate the primary methods for generating adversarial samples. It should be noted that recent research has shown the limitations of some earlier methods that are still listed here for reference even though more effective methods have been introduced.
 
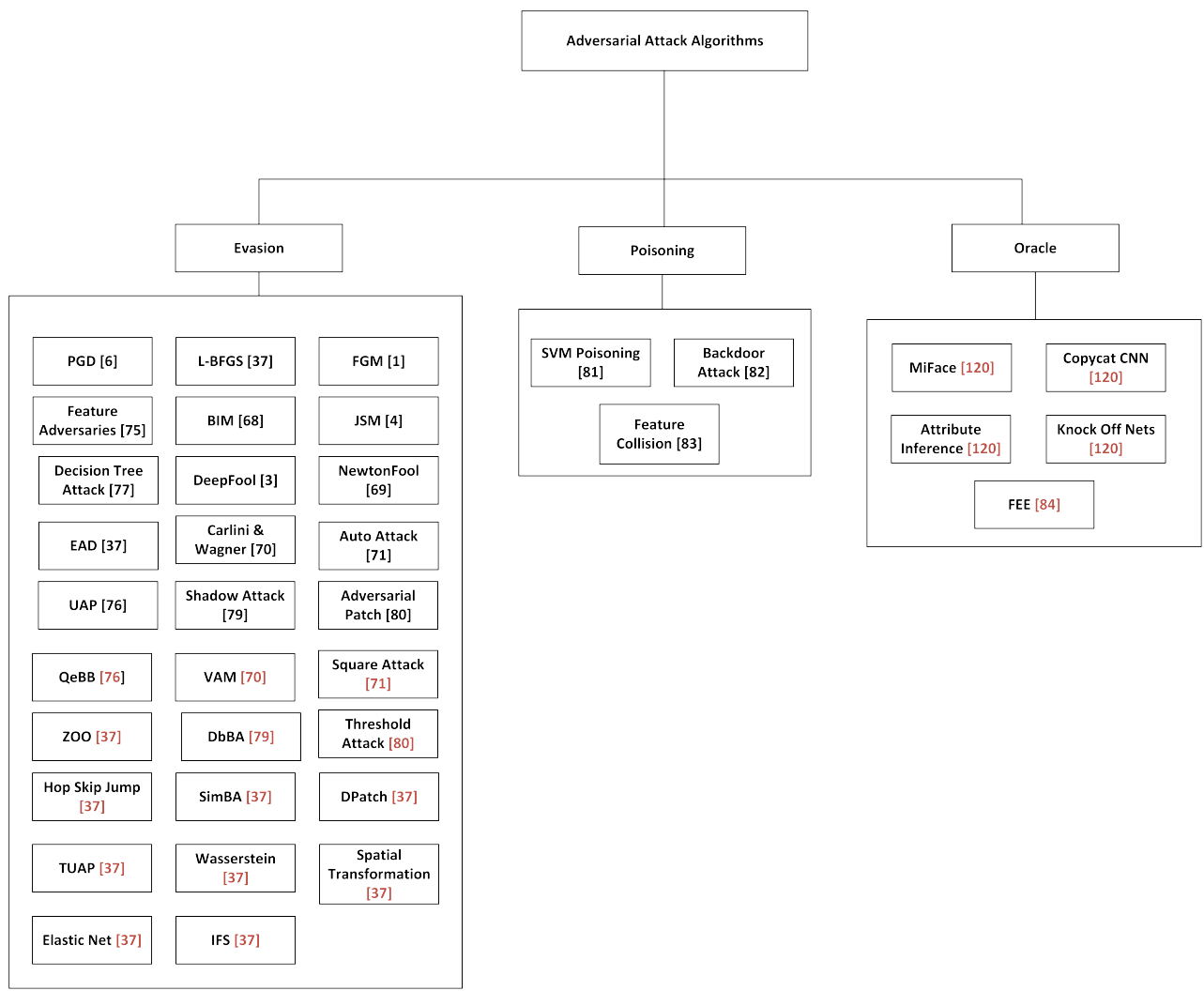
\begin{figure*}[h!]
   \centering
    \begin{tikzpicture}[
      >=latex',
      auto
    ]
      \node [intg] (kp)  {Adversarial Attack Algorithms};
      \node [int]  (ki1) [node distance=1.5cm and -1cm,below left=1.5cm and 0.05cm of kp] {Poisoning};
      \node [int]  (ki2) [node distance=1.5cm and -1cm,below=of kp] {Evasion};
      \node [int]  (ki3) [node distance=1.5cm and -1cm,below right=1.5cm and 0.05cm of kp] {Oracle};
  \matrix [draw,column sep={0.1cm},nodes=draw,below=of ki3] (Oracle)
  {
    \node { Mi-Face\cite{fredrikson2015model}}; & \\
    \node {Copycat CNN\cite{correia2018copycat}} ;    \\
    \node    {Attribute inference\cite{fredrikson2015model}};  &\\
    \node     {KnockOff Nets\cite{orekondy2019knockoff}};   \\
    \node    { FEE\cite{jagielski2020high}};   &   \\
  };

  \matrix [draw,column sep={0.1cm},nodes=draw,below=of ki1] (Poisoning)
  {
    \node(a) { SVM Poisoning\cite{biggio2012poisoning}}; &   \\
    \node    {Backdoor Attack\cite{gu2017badnets}};  & \\
    \node    { Feature Collision\cite{shafahi2018poison}};   &   \\
  };
 \matrix [draw,column sep={0.1cm},nodes=draw,node distance=4.2cm and -1cm,below=of ki2] (Evasion)
  {
    \node { PGD\cite{madry2017towards}}; & \node {L-BFGS\cite{szegedy2013intriguing} } ;& \node {FGM\cite{goodfellow6572explaining} } ;    \\
     \node { Feature Adversaries\cite{sabour2015adversarial}}; & \node {BIM\cite{kurakin2016adversarial} } ;& \node {JSM\cite{papernot2016limitations}} ;    \\
    \node { Decision Tree Attack\cite{papernot2016transferability}}; & \node {DeepFool\cite{moosavi2016deepfool}} ;& \node {NewtonFool\cite{jang2017objective} } ;    \\
     \node { EAD\cite{chen2018ead}}; & \node {Carlini and Wagner\cite{carlini2017towards}} ;& \node {AutoAttack\cite{croce2020reliable}} ;    \\
    \node { UAP\cite{moosavi2017universal} }; & \node {Shadow Attack\cite{ghiasi2020breaking}} ;& \node {Adversarial Patch\cite{brown2017adversarial}} ;    \\   
    \node {Square Attack\cite{croce2019provable}} ;    
    &\node { ZOO\cite{chen2018ead}}; & \node {DbBA\cite{ghiasi2020breaking}} ;\\ \node {Threshold Attack\cite{brown2017adversarial}} ;    
       &\node { Hop Skip Jump\cite{chen2018ead}}; & \node {SimBA\cite{chen2018ead}} ; \\ \node {DPatch\cite{chen2018ead}} ;    
      & \node { TUAP\cite{chen2018ead}}; & \node {Wasserstein\cite{chen2018ead}} ;\\ \node {Spatial Transformation\cite{chen2018ead}} ;   & 
    \node    { Elastic Net\cite{chen2018ead}}; & \node {IFS\cite{chen2018ead}} ;   &   \\
  };
      \draw[-] (kp) -- ($(kp.south)+(0,-0.75)$) -| (ki1) node[above,pos=0.25] {} ;
      \draw[-] (kp) -- ($(kp.south)+(0,-0.75)$) -| (ki2) node[above,pos=1.25] {};
      \draw[-] (kp) -- ($(kp.south)+(0,-0.75)$) -| (ki3) node[above,pos=0.25] {};
     \draw[-] (ki3) -- ($(ki3.south)+(0,-0.75)$) -| (Oracle) node[above,pos=0.25  ] {};
      \draw[-] (ki2) -- ($(ki2.south)+(0,-0.75)$) -| (Evasion) node[above,pos=0.25  ] {};
      \draw[-] (ki1) -- ($(ki1.south)+(0,-0.75)$) -| (Poisoning) node[above,pos=0.25  ] {};
    \end{tikzpicture}
    \caption{Adversarial Attack Algorithms}
    \label{fig:adv_attack_algo}
  \end{figure*}

In the previous section \ref{threat_model}, we described our threat model for adversarial attacks in network security. In this section, we introduce a classification method for the various adversarial attack algorithms. As seen in Figure \ref{fig:adv_attack_algo} our classification method is based on the adversary strategy described in section \ref{strategy}.

\subsubsection{Evasion Attacks}
Evasion attacks attempt to mislead the machine learning system during the testing or inference phase. Below we highlight adversarial attack methods that fall within this category of evasion attacks. The attacks are further divided into \textbf{Gradient-based} and \textbf{Gradient-free} attacks.

\begin{itemize}

\item Gradient-based attacks: Szegedy et al. \cite{szegedy2013intriguing} studied how adversarial samples could be generated against neural networks for image classification. The L-BFGS (Limited Broyden-Fletcher- Goldfarb-Shanno) method was then introduced, which used an expensive linear search method to find the optimal values of the adversarial samples. In a different approach proposed by Goodfellow et al. \cite{goodfellow6572explaining} called the Fast Gradient Sign Method (FGSM), adversarial samples are created by finding the maximal direction of positive change in the loss. This is a faster method than the L-BFGS method since only a one-step gradient update is performed along the direction of the sign gradient at each level. A major limitation of the Fast Gradient Sign Method and similar attack methods is that they work based on the assumption that the adversarial samples can be fed directly into the machine learning model. This is far from being practical since most attackers would seek to access the machine learning models through devices such as sensors \cite{yuan2017adversarial}. The Basic Iterative Method (BIM) proposed in \cite{kurakin2016adversarial} overcomes this limitation by running the gradient update in multiple iterations.

The Jacobian-based Saliency Map Attack (JSMA) was introduced by Papernot et al. \cite{papernot2016limitations}. For the attack, the Jacobian matrix of a given sample is computed to find the input features of that sample which most significantly impacts the output. Subsequently, a small perturbation is created based on that input feature for generating the adversarial attack. DeepFool was proposed by Moosavi et al. \cite{moosavi2016deepfool} as a method for creating adversarial samples by finding out the closest distance between original input and the decision boundary for adversarial samples. They were able to determine that by using a related classifier, the closest distance which would correspond to the minimal perturbation for creating an adversarial sample will be the distance to the hyperplane of the related classifier.

Jang et al. \cite{jang2017objective} presented the NewtonFool attack, an algorithm that is based on gradient-descent to find adversarial samples. This attack is similar to Deepfool \cite{moosavi2016deepfool} but more effective in producing good adversarial samples and reduces the confidence probability of the correct class. They exploit the softmax layer and control the step size and how small the perturbation could be. Carlini et al. \cite{carlini2017towards} developed Carlini and Wagner Attack, a targeted attack specifically for existing adversarial defense methods. It was discovered that defenses such as defensive distillation \cite{papernot2016distillation} were ineffective towards the Carlini and Wagner attack. Madry et al. \cite{madry2017towards} proposed the Projected Gradient Descent (PGD) adversarial attacks that is more robust than FGSM. This form of attack utilizes a multi-step approach with a negative loss function. It overcomes the network overfit problem, and shortly comes of FGSM adversarial samples. It is more robust than FGSM, which utilizes the first-order network information, and it works well in large-scale constraints.  In $\ell_\infty$-ball, PGD iterate to explore the maximum loss. 

Croce et al. \cite{croce2020reliable} proposed Auto Attack, an attack that overcomes and remedy the weaknesses of Projected Gradient Descent (PGD) \cite{madry2017towards} that lead to model robustness false outcomes. First PGD attack use fixed step size with cross-entropy as a loss function that causes the failure as identity by \cite{mosbach2018logit}. In \cite{croce2020reliable}, they use a new gradient-based scheme without step size selection with different loss function. With these two changes, two versions of PGD produced with free parameters in the number of iteration. They also integrate the new PGD versions with FAB-attack \cite{croce2019minimally} and Square attack \cite{croce2019provable} to produce a parameter-free attack called AutoAttack. The authors also integrated two Auto Attack and were tested on a large scale on 40 classifiers.

Sabour et al. \cite{sabour2015adversarial} proposed a new adversarial image attack that not only focus on the class label but in the internal representations. The attack, known as Feature Adversaries enables the possibility to deceive a trained DNN to mystify any source image with other target image by finding a small perturbation from the source image that create similar internal representation to the target image and not related to the source image. The authors however take into consideration  that such adversaries are not outliers. Universal Perturbation \cite{moosavi2017universal} was proposed by Moosavi et al. as an algorithm to calculate a universal small image perturbation to misclassify a state-of-the-art deep neural network classifier. The main focus of this algorithm was to find the perturbation vector that deceives classifier on all data point samples. This fix perturbation is existed to lead changes in image label gradually to build the universal perturbation. 


\item Gradient-free Attacks: Decision Tree Attack was proposed by Papernot et al. \cite{papernot2016transferability} this type of black-box attacks use transferability of adversarial samples between and within different classifiers, including Deep neural network. Logistic regression, decision trees, support vector machines (SVM), ensembles, and nearest neighbors. They demonstrated that black box attacks are feasible to a machine learning algorithm that not using deep neural networks and adversarial samples works well between and across models using the same and different machine learning techniques. Chen et al. \cite{chen2018ead} proposed an adversarial attack algorithm to attack DNN based on elastic-net regularization in feature $L_1$ and $L_2$  called elastic-net attacks to DNNs (EAD).  EAD considers state-of-the-are $L_2$ and $L_infinty$ Authors demonstrated that EAD could break undefended and defensively distilled DNNs. They also improve the transferability of attacks and adversarial training.  Shadow Attack was proposed by Ghiasi et al. \cite{ghiasi2020breaking} which is a new method for attacking systems that rely on certificates and fool certified robust networks to assign the wrong label to an image and produce a spoofed secure robustness certificate for the adversarial example. Adversarial Patch, proposed by Brown et al. \cite{brown2017adversarial} present universal, robust, and targeted adversarial patches for the real world that do not require any knowledge about what image they are attacking.  Those adversarial samples can be used to attack any classifier, and they work with many transformations that exist defense methods may not be robust to such a massive transformation. The adversarial patch leads the classifier to switch class labels to any target class. Chen et al. \cite{chen2019hopskipjumpattack} develop HopSkipJumpAttack based on a decision-based attack that is a type pf black-box attack. This algorithm generates iterative targeted and untargeted adversarial samples with minimum distance. This attack demonstrates superior efficiency over various state-of-the-art decision-based attacks. The iteration in the algorithm is based on gradient direction, step size, and boundary search. 
 
\end{itemize}

\subsubsection{Poisoning Attacks} A poisoning attack also known as causative attack, uses direct or indirect means to alter the data or the model. Poisoning attacks occurs either by injecting false data, manipulating the original data, or corrupting the model logic.

\begin{itemize}

    \item Data Injection: Biggio et al. \cite{biggio2012poisoning} proposed a gradient ascent based attack based on SVM that attacks the input data that lead to maximize the non-convex surface error and increase classifier classification at the test time. Gu et al. \cite{gu2017badnets} proposed BadNets, which perform adversarial attacks by discovering the backdoored neural network or BadNet. The attack is based on a full or partial outsourced training process where attacker provides the user with a trained model with a backdoor that causes a targeted misclassification and degrade in the accuracy in some cases called backdoor trigger. For example, in autonomous driving,  an attacker provides the user with a street sign detector that is backdoored, which classify stop sign well in most cases except when the stop signs have a particular sticker in classifying it as speed limit signs. This type of attack occurs under two scenarios user outsource trained model or download a pre-trained model.
    
    \item Data Manipulation: Feature Collision Attack proposed by Shafahi et al. \cite{shafahi2018poison} presents a watermarking poisoning attack based on optimization-based to craft a clean label attack to target the behavior of a neural network classifier on a specific instance. This attack uses enhanced preservation techniques to make it difficult to be detected. 
    
\end{itemize}

\subsubsection{Oracle Attacks}
In an oracle type adversarial attack, an adversary who has been given a oracle prediction access to a model, steals a copy of a remotely deployed machine learning model. This enables the adversary to duplicate the functionality of the model, i.e "steal the model" \cite{tramer2016stealing}. This attack has become increasingly common due to the increase in Machine Learning as a Service "MLaaS" offerings where several companies that offer cloud-based Machine Learning services e.g. Google, Amazon, and BigML, provide easy-to-use web APIs to manage client interaction.  
\begin{itemize}

\item Inversion Attacks: Fredrikson et al. \cite{fredrikson2015model} exposed the privacy issues with providing access to machine learning API. Their study demonstrated how an adversary could utilize the confidence information of a model to result in model inversion attacks. The attack, which is implemented as a function called MI-Face attack, enables an adversary to extract pictures of subjects from a trained machine learning model. 

\item Inference Attacks: Fredrikson et al. \cite{fredrikson2015model} proposed the attribute inference attack which could be launched either as a white-box or black-box attack.

\item Extraction Attacks: Correia-Silva et al. \cite{correia2018copycat} demonstrated how an adversary could create a substitute model from a black-box convolutional neural network (CNN) model by querying the black-box model with random non-labeled data. A more intriguing aspect of this oracle type of extraction attack is the fact that dataset used to persuade the model was not related to original problem domain. Orekondy et al. \cite{orekondy2019knockoff} proposed Knockoff Nets which are capable of stealing the functionality of a fully trained model using a two-step approach. The adversary first obtains predictions from the model by querying a set of input data, then the data-prediction pairs are used to create a substitute model known as a "knock-off" model. Their approach uses a reinforcement learning approach with demonstrated query efficiency and performance gains, compared to other oracle type attacks. Jagielski et al \cite{jagielski2020high} proposed the Functionally Equivalent Extraction (FEE) attacks which explore accuracy and fidelity objectives within the space of model extraction by improving the query efficiency of learning attacks. Their method is demonstrated to be practical for high parameter models in the range of millions. In their attack method, an adversarial model is produced whose architecture and weights are identical to the oracle. 

\end{itemize}

\section{Adversarial Attack Classification}

Multiple studies \cite{das2018taxonomy} \cite{hansman2005taxonomy} have sought to differentiate the different domains of network security into multiple fragmented domains. A common approach for example make attempts at differentiating malware and spam detection from intrusion detection \cite{rosenberg2021adversarial}. We find that this attempt of fine grained classification results in redundancy, since the task of malware or phishing detection in a network could be considered an intrusion detection task. As such, in this survey, we consider cyber attacks against a network as an attempt by an adversary to intrude the network with a malicious payload. We identify malicious payload in a network to consist of three broad types: malicious files (malware), malicious text (spam) and malicious url links (phishing). We note that attackers may use a combination of all three payloads in most cyber attacks. For example, a spam email may also contain a link to a malicious url or contain a malicious file attachment. This payload approach becomes even more crucial in our study on adversarial attacks within the network security domain. We realise from our study that this distinction plays an important role in providing an accurate classification of adversarial attacks within the network security domain, as compared to other domains such as computer vision.

\begin{figure}[thpb]
    \centering
    \includegraphics[width=0.85\linewidth,keepaspectratio=true]{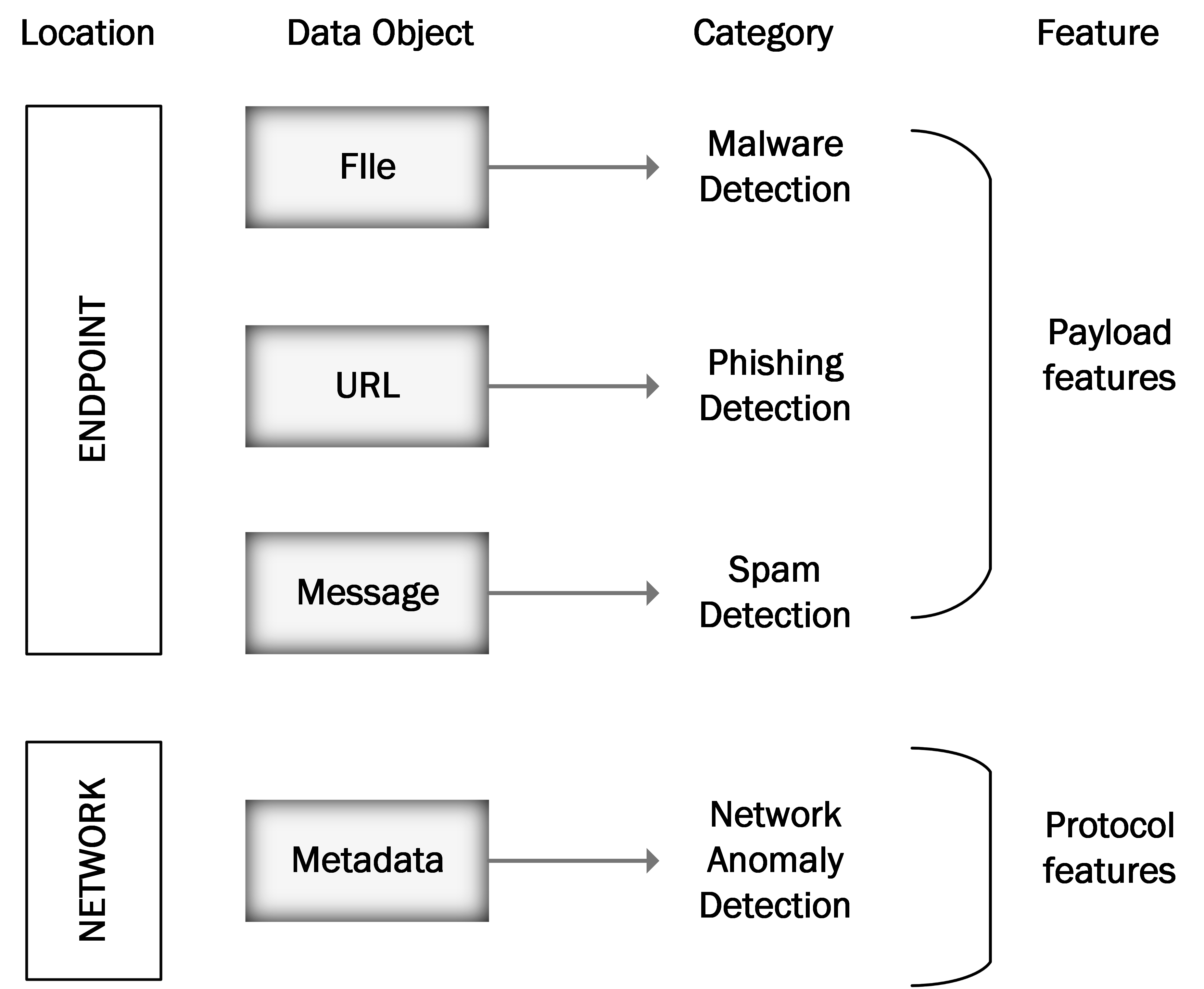}
    \caption{Adversarial attack classification}
    \label{fig:attack_classification_old}
\end{figure}

In this section, we introduce a classification method for adversarial attacks in network security based on network security task. Our classification approach considers the data object which is being manipulated by the adversary. The feature scope of the adversarial attack corresponds to the data object as shown in Figure \ref{fig:attack_classification_old}. 

For the scope of this study, we consider adversarial attacks based on the actual payload which is being attacked in context. When a message is being transmitted from a sender to a receiver, the payload represents the portion of the transmitted data that is actually the intended message. For example, when an email is sent, the payload consist of the message body, attachments, and URL links. Headers and metadata which help to facilitate the delivery of the payload are not considered as part of the payload, within the context of our study. Hence, the protocol overhead is not considered as part of the actual data. 

Our approach for classifying adversarial attacks in network security is based off this approach, as shown in Figure \ref{fig:attack_classification_old}. This is known as feature scope based classification, which refers to what features are being manipulated or perturbed by the adversary in other to generate an adversarial sample. Adversarial attacks against malware detection, phishing detection and spam detection applications try to perturb the payload features such as a binary file, a URL, or an email message. These attacks are categorized as adversarial attacks against endpoint protection systems. Conversely, we also have adversarial attacks against network anomaly detection applications and these type of attacks will seek to perturb protocol features such as the network metadata or protocol headers. We categorize these attacks as adversarial attacks against network protection systems.

Network security domain that utilize machine learning techniques fall into four broad categories namely malware detection, phishing detection, spam detection and network anomaly detection. We illustrate this categorization in Figure \ref{fig:attack_classification_old}. The first three categories of network security tasks are considered as endpoint based protection. Machine learning applications within this endpoint based protection category are typically initiated with payload features. Network protection primarily constitutes network anomaly detection and machine learning applications within this category are typically initiated with protocol features. Our study only considers active attacks against a network, and passive attacks such as eavesdropping are not within the scope of this study. Adversarial attacks hence seek to generate adversarial samples using specific data objects.

In contrast to adversarial attacks in the field of image processing or computer vision, network security's adversarial learning is more challenging. This occurs because even very slight modifications to URLs, spam, packets, or malware bytes of the binary files can significantly alter the functionality of the data. In computer vision, the addition of tiny perturbations to an image sample does not alter the human perception of the image and same as in speech processing. Text processing and network security filtering techniques are similar in this regard since a very slight change in the input such as a word or a byte will alter the meaning of the text or the data functionality. Hence, approaches for generating adversarial samples in the domain of machine learning-based network security filtering systems need to occur in such a way that the malicious functionality is not distorted. Several approaches for achieving these adversarial attacks have been researched and are discussed in the sections below.

\subsection{Adversarial attacks against Malware Detection}

A major component of endpoint protection in network security is malware detection. Yet, malware detection remains a challenging problem in network security. Between 2009 and 2019, the number of new malware digital signatures has increased by over 2000 percent \cite{avtest2019}. Therefore, traditional malware detection systems that rely solely on digital signatures have become less effective. Significant effort has been made in the use of machine learning to protect against malware attacks. Several researches have shown the vulnerability of these machine learning models to adversarial attacks. The most common approach is the addition of selected sequence of bytes to the binary file. Several approaches have been considered for synthesizing this sequence of bytes as discussed below. 


Malware detection may be based on static analysis, in which the malware is detected without executing the code. Alternatively, dynamic analysis for malware detection typically executes a suspicious malware sample in a sandbox in an attempt to discover dynamic behavioural patterns such as API call sequences.

\subsubsection{Iagodroid}
One of the earliest attacks against machine learning based malware detection systems was the Iagodroid attack \cite{calleja2018picking}. Iagodroid uses a method to induce mislabelling of malware families during the triaging process of malware samples. Their evasion rate reached 97 percent.

\subsubsection{Stingray}
Suciu et al \cite{suciu2018does} proposed an adversarial attack against malware using the 'FAIL' model. Their study focuses on constraints of obscurity and transferability in order to realize a targeted poisoning attack. StingRay succeeded in half of the test cases.

\subsubsection{Texture Perturbation Attacks}
Researchers have deployed visualization techniques similar to computer vision and adapted it for malware classification \cite{han2015malware}. This involves conversion of malware binary code into image data. The Adversarial Texture Malware Perturbation Attack  (ATMPA) achieved a  100 percent effectiveness in defeating visualization based machine learning malware detection system and also resulted in 88.7 percent transfer-ability rate \cite{liu2018adversarial}.  The attack model for ATMPA works by allowing the attacker to distort the malware image data during the visualization process.

\subsubsection{Android malware attack in Problem space} \cite{pierazzi2020intriguing} et al formalized an approach for problem space adversarial evasion attacks against machine learning based android malware detection systems. Their
study identified four main contraints which are characteristic of any problem space attack. Their study adopted a technique which automates the generation thousands of realistic and inconspicuous adversarial malware samples, further buttressing the notion of adversarial malware as a service as a real threat in network security. Their attack led to a misclassification rate of 100.0 percent on the successfully generated samples.

\subsubsection{EvadeDroid}
Bostani et al. \cite{bostani2021evadedroid} presented EvadeDroid, another problem space Android evasion attack. EvadeDroid is a query-efficient black-box attack, that can fool ML-based Android malware detectors without altering the functionality of the original malware samples. It uses an n-gram-based similarity method to select candidate donors for gadget extraction to change malware samples into benign ones through an iterative and incremental manipulation technique. Their experimental results demonstrated that EvadeDroid's evasion rates are 81, 73, 75, and 79 percent for DREBIN, Sec-SVM, MaMaDroid, and ADE-MA, respectively. 

\subsubsection{EvnAttack}
EvnAttack is an evasion attack model that was proposed in \cite{chen2017adversarial} which manipulates an optimal portion of the features of a malware executable file in a bi-directional way such that the malware is able to evade detection from a machine learning model based on the observation that the API calls differently contribute to the classification of malware and benign files. The detection model's false negative ratio almost reached 1 (100 percent), which means almost all malware samples are misclassified.

\subsubsection{AdvAttack}
AdvAttack was proposed in \cite{chen2017secmd} as a novel attack method to evade detection with the adversarial cost as low as possible. This is achieved by manipulating the API calls by injecting more of those features which are most relevant to benign files and removing those features with higher relevance scores to malware. AdvAttack increased the classifier's false negative ratio to 71 percent while degrade the accuracy of the classifier to 58.5 percent.

\subsubsection{MalGAN}
To combat the limitations of traditional gradient-based adversarial sample generation, the use of a generative adversarial network (GAN) based algorithm for generating adversarial samples has been proposed. Generative models have been mostly used for input reconstruction by encoding an original image into a lower-dimensional latent representation  \cite{kos2018adversarial}.  The latent representation of the original input can be used to distort the initial input to create an adversarial sample. MalGAN proposed by \cite{hu2017generating} leverages on generative modeling techniques to evade black-box malware detection systems with a detection rate close to zero.

\subsubsection{GAPGAN}
Yuan et al. \cite{yuan2020black} introduced GAPGAN, an adversarial attack framework that generates adversarial examples against binaries-based malware detection through GANs. Adversarial perturbations are appended to the original malware binaries to maintain its malicious functionality. They tested GAPGAN on deep learning and MalConv detectors. GAPGAN's success rate reached 100 percent attack with appending payloads of 2.5 percent of the total length of the original data.

\subsubsection{Black-Box Attacks against RNN Based Malware Detection Algorithms} Hu et al. \cite{hu2018black} implemented a generative recurrent neural network (RNN) which generates sequential adversarial samples. In their study, the Gumbel-Softmax approach is used to approximate generated discrete API's. Before their attack, the victim's RNN malware detection rates ranged from 90.74 to 93.87 percent. After their adversarial attack, the detection rates on adversarial examples ranged from 0.44  to 3.03 percent.

\subsubsection{Adversarial Deep Learning for Robust Detection of Binary Encoded Malware}
Al-Dujaili et al \cite{al2018adversarial} proposed a method of generating adversarial malware samples with a focus on preserving the malicious functionality of the binary encoded files. They also introduce a mitigation framework known as SLEIPNIR which employs the saddle-point optimization technique to learn malware detection models.

\subsubsection{Deceiving End-to-End Deep Learning Malware Detectors using Adversarial Examples}
The authors Kreuk et al. \cite{kreuk2018deceiving} introduced a novel approach for creating adversarial malware samples by injecting a small sequence of bytes to the binary file. The approach was also found to be transferable across different malware files and families. In their study, they evaluated the effectiveness of adversarial malware samples based on five metrics namely (1) File transferability, (2) Spatial Invariance (3) payload size, (4) entropy (5) Functionality preservation. Their study was based on only white box attacks and was not evaluated as white box scenarios. Their injection procedure resulted in an evasion rate of 99.21 and 98.83 percent.

\subsubsection{Adversarial Examples on Discrete Sequences for Beating Whole-Binary Malware Detection}
The authors \cite{kreuk2018adversarial} focus on adversarial attacks against Convolutional Neural Network (CNN) based end to end malware detectors. End to end malware detectors such as Malconv \cite{raff2018malware} function quite different from most deep learning based malware detectors in the sense that they take the whole malware binary file as an input. To achieve their aim, a loss function was which functions as a surrogate loss function proposed which enforces the modifications in the embedding space. Thus, the authors were able to modify the embedding vector in order to reconstruct the modified binary, which becomes the adversarial malware sample. To preserve the functionality of the malware binary, a unique section of payload bytes is perturbed and appended to the original malware binary file instead of perturbing the original binary file. Thus by adding perturbations in the embedding vector space and reconstructing new binary files from the adversarial example. This attack's evasion rate reached 100 percent.

\subsubsection{Adversarial-Example Attacks Toward Android Malware Detection System}. 
MalGAN \cite{hu2017generating} proposed a black-box adversarial-example attacks toward Android malware detection, in which adversarial examples are generated using a generative adversarial network (GAN) without requiring the knowledge about the target. Unfortunately, the effectiveness of Malgan is affected, if a firewall is incorporated into the malware detection system. Adversarial attacks were also studied against cloud-based Android malware detection systems. Li et al. proposed a bi-objective GAN type adversarial attack against android malware detection systems. Their technique has the novelty of implementing a GAN with two discriminators in which one discriminator contends against the firewall while the other discriminator contends against the malware detector. This study was the first study to target a firewall-equipped Android malware detection system.

\subsubsection{Adversarial malware sample generation method based on the prototype of deep learning detector}
Qiaoa et al.\cite{qiao2022adversarial} presented a method for generating adversarial malware to fool the deep learning-based malware detection systems. The post-hoc interpretability of deep learning is used by the authors to direct the malware file's updates. Based on their experiments, the time to generate their adversarial malware is less than other attacks. The fooling rate of this attack reached 92 percent.

\subsubsection{Slack Attacks}
A byte-based convolutional neural network (MalConv) was introduced by Raff et al. \cite{raff2017malware}. Unlike image perturbation attacks \cite{szegedy2013intriguing}, where the fidelity of the image is of little concern, attacks that alter the binaries of malware files must maintain the semantic fidelity of the original file because altering the bytes of the malware arbitrarily could affect the malicious effect of the malware. This problem could be solved by appending adversarial noise to the end of the binary \cite{kolosnjaji2018adversarial}. This prevents the added noise from affecting the malware functionality. The Random Append attack and Gradient Append attacks are two types of append attacks which work by appending byte values from a uniform distribution sample and gradually modifying the appended byte values using the input gradient value. Two additional variations of append attacks; the benign append and the FGM Append were introduced by Suciu et al. \cite{suciu2018exploring} which improves the long convergence time experienced in previous attacks. When malware binaries have exceeded the model's maximum size, it is impossible to append additional bytes to them. Hence a slack attack proposed by Suciu et al. \cite{suciu2018exploring} exploits the existing bytes of the malware binaries. The most common form of the slack attack is the Slack FGM Attack which defines a set of slack bytes that can be freely modified without breaking the malware functionality.

\subsubsection{Attack and Defense of Dynamic Analysis-Based, Adversarial Neural Malware Detection Models}

Stokes et. al \cite{stokes2018attack} proposed adversarial attacks against dynamic analysis-based malware detection systems. Their work focuses on  different strategies of crafting adversarial samples for deep learning based dynamic analysis of malware samples. Their study is motivated in the fact that static analysis based deep learning malware classifiers only classify the content of the unknown file without execution, and become less effective when faced with packed or encrypted malware files. In addition, they propose a defense mechanism known as the weight defense mechanism.  The compare their defence technique to existing defenses such as distillation and ensemble defenses. They however did not compare their study to the more popular approach of adversarial training, which is a proven method for reducing the vulnerability  deep learning classifiers to adversarial samples. Their study also indicates that adding more hidden layers to the neural network significantly improves the robustness of the deep learning based malware classifier to adversarial samples.

\subsection{Adversarial attacks on Spam Detection}

Spam detection is a significant endpoint protection component, used to protect users from unsolicited digital communications. Machine learning techniques are widely used for current spam filtering applications, most of which utilize supervised learning methods \cite{crawford2015survey}. Multiple adversarial attacks on machine learning-based spam detection systems are discussed below.

\subsubsection{Adversarial classification}
Dalvi et al \cite{dalvi2004adversarial} were the first to introduce a formal framework with corresponding algorithms to describe the problem of adversarial attacks against machine learning based spam detectors. In their study, they seek the minimum cost camouflage (MCC) of a data sample \(x\) to generate an adversarial sample MCC(x) with the minimum cost, for which the classifier outputs a negative sample. Similar studies \cite{lowd2005good} had considered adversarial attacks against spam detectors albeit not machine learning based.

\subsubsection{Attacks on Statistical Spam filters}
Several spam filters such as SpamAssasin, SpamBayes, Bogofilter are based on the popular Naive Bayes Machine learning algorithm which was first applied to filtering junk email in 1998 \cite{sahami1998bayesian}. A variety of good word attacks introduced by Lowd \cite{lowd2005good} were successfully evading the machine learning models from detecting spam or junk emails. Using these attacks, an attacker can get 50 percent of currently blocked spam past a typical spam ﬁlter.

\subsubsection{Exploiting Machine Learning to Subvert Your Spam Filter}
Nelson et al. \cite{nelson2008exploiting} showed in 2008 that an attacker could effectively disable the SpamBayes spam filter with small information and little control over training data. Their introduced Usenet dictionary poisoning attack caused misclassification of 36 percent of ham messages with only 1 percent control over the training data. They have also presented a new class of focused attacks that stop victims from receiving specific email messages. With knowledge of only 30 percent of the target’s tokens, their focused attack altered the classification of the target email 60 percent of the time.

\subsubsection{Attacks against crowd-turfing detection systems}
Machine learning techniques are used to identify misbehavior includes fake users in social networks and detect users who pays for sites to have fake accounts. Malicious crowdsourcing or crowd-turfing systems are used to connect users who are willing to pay, with workers who carry out malicious activities such as generation and distribution of fake news, or malicious political campaigns. Machine learning models have been used to detect crowdturfing activity with up to 95 percent accuracy particularly in detecting the accounts of crowdturfing workers \cite{wang2014man}. However, malicious crowdsourcing detection systems are highly vulnerable to adversarial evasion and poisoning attacks.

\subsubsection{Attacks Against ML for Keystroke Dynamics}
Negi et al. \cite{negi2017adversarial} created adversarial keystroke samples that misled an otherwise accurate classifier into accepting the artificially generated keystroke samples as belonging to an authentic user. Almost 50 percent of the tested users were compromised after their attack.

\subsubsection{Attacks against ML for credit card fraud detection}
Zeager et al. \cite{zeager2017adversarial} examined how a logistic regression classifier used as a fraud detection mechanism, could be adversarially attacked to cause a number of fraudulent transactions to go undetected. Previous studies have similar models which are based on game theory to investigate adversarial attacks against credit card fraud detection and email spam detectors. However, the authors introduced a new framework which successfully produced an improved AUC score on multiple iterations of the validation sets compared to the performance of the models which credit card companies had previously used.

\subsubsection{Crafting Adversarial Email Content against Machine Learning Based Spam Email Detection}
Wang et al. \cite{wang2021crafting} proposed two methods to create adversarial email content to bypass spam detectors. The first approach approximates the Term Frequency–Inverse Document Frequency) TF-IDF values in the resultant adversarial examples and the second method recognizes and adds a group of significant words to fool the detectors. They tested their work on multiple machine language models like; KNN, SVM, decision tree, and logistic regression, in both white-box and black-box attack scenarios. Their attacks' success rates ranged from 2.2 to 98.9 percent, which is inconclusive. However, they concluded that the second method is more effective.

\subsubsection{Marginal Attacks of Generating Adversarial Examples for Spam Filtering}
Zhaiquan et al. \cite{zhaoquan2021marginal} created the marginal attack, which generates adversarial samples that can deceive naive bayesian spam filters by selecting sensitive words from a sentence and then add them at the end of the sentence. Their experiments showed that adding just one word to the message could reduce the model's accuracy from 93.6 to 55.8 percent. They also tested the transferability of the generated adversarial samples against standard machine learning filters like logic regression, decision tree, and linear support vector. In some cases, the accuracy of these filters could drop from 100 to 1.5 percent.

\subsubsection{Universal Adversarial Perturbations and Image Spam Classifiers}
Phung et al.\cite{phung2021universal} evaluated numerous adversarial attack methods against deep learning-based image spam classifiers, and they found that the universal perturbation method is the most harmful. So they used this approach to create a novel transformation-based adversarial attack that was capable of creating tailored “natural perturbations” in image spam.  In some cases, their suggested attack can lower the model’s accuracy to reach 23.7 percent.

\subsection{Adversarial attacks against Phishing Detection}

Phishing detection is a critical endpoint protection element aimed to save the users from serious fraudulent actions like; money stealing and accessing private information. There are multiple techniques for phishing detection like \cite{shahrivari2020phishing}; List-base approach, Visual similarity-base approach, and Heuristics and machine learning-based approach, which is the most popular method now. Several adversarial attacks on machine learning-based phishing detection systems are discussed below.

\subsubsection{FIGA}

Gressel et al.\cite{gressel2021feature} proposed the Feature Importance Guided Attack (FIGA) to fool phishing detection models by perturbing the most effective features of the input in the direction of the target class. It is a model-agnostic gray-box attack that needs knowledge of the feature representation of the victim model. FIGA was tested on eight different phishing detection models, and it reduced the F1-score of the models from 0.96 to 0.41 on average.

\subsubsection{Bypassing Detection of URL-based Phishing Attacks Using Generative Adversarial Deep Neural Networks}

AlEroud et al. \cite{aleroud2020bypassing} presented an evasion technique that attacks URL phishing detection systems via Generative Adversarial Networks (GAN). Their generated samples can deceive Blackbox phishing detectors even when those detectors are created using refined methods like those relying on intra-URL similarities. Their experiments revealed that some classifiers were unable to identify any of the adversarial examples leading to zero true positive rates. At the same time, the false positive rates are increased, which indicates the percentage of benign examples classified as phishing.

\subsubsection{Generating Optimal Attack Paths in Generative Adversarial Phishing}

Al-Qurashi et al. \cite{al2021generating} proposed a method that creates adversarial phishing attacks by discovering optimal subsets of features that lead to a higher evasion rate. To achieve this, multiple feature engineering techniques are used, such as Recursive Feature Elimination, Lasso, and Cancel Out. Their experiments revealed that their attack has better evasion capability than Generative Adversarial Deep Neural Network (GAN) which randomly perturbs features.

\subsubsection{Advanced evasion attacks and mitigations on practical ML‐based phishing website classifiers}

Song et al. \cite{song2021advanced} introduced multiple mutation-based techniques, differing in the knowledge of the target classifier (white, gray, and black boxes). They also proposed a sample‐based collision attack to acquire the knowledge of the target model, in the cases of white- and gray-box scenarios. Their evasion attacks fooled the classifiers without changing the functionalities and appearance of the samples. Their attack's success rate varied depending on the knowledge and the attacked model. Attacks on Google's phishing page filter achieved a 100 percent attack success rate. Their transferability attack on BitDe-fender's industrial phishing page classifier, TrafficLight, achieved 81.25 and 50 percent transferability attack rates in the black‐ and gray‐box scenarios.

\subsection{Adversarial attacks against Network Anomaly Detection}

Network anomaly detection devices learn network activity patterns and detect irregularities. They must continuously scan the network, analyze encrypted data, and spot anomalies in real-time. Machine learning ticks all these boxes, that's why it is used extensively in modern Network anomaly detection tools, however, researches have found some ways to attack them. Multiple of these adversarial attacks are discussed below.

\subsubsection{IDSGAN}
IDSGAN was proposed by Lin et al. \cite{lin2018idsgan} for generating adversarial attacks targeted towards intrusion detection systems. IDSGAN is based on the Wasserstein GAN \cite{arjovsky2017wasserstein} which uses a generator, discriminator and a black-box. The discriminator is used to imitate the black-box intrusion detection system and at the same time provide the malicious traffic samples. IDSGAN can lower the detection rates of some IDS models to approximately zero percent. 

\subsubsection{TCP Obfuscation Techniques}
Another method for evading machine learning based intrusion detection systems is the use of obfuscation techniques. Homolial et al. \cite{homoliak2018improving} proposed the modification of various properties of network connections to obfuscate a TCP communication which successfully evades a wide variety of intrusion detection classifiers.

\subsubsection{Deep Adversarial Learning in Intrusion Detection: A Data Augmentation Enhanced Framework}
Zhang el al. \cite{zhang2019deep2} proposed a framework which incorporates deep adversarial learning with statistical learning in a manner which exploits learning-based data-augmentation. In the study, the Poisson-Gamma joint probabilistic generative model is used to synthesize adversarial samples.

\subsubsection{Generative Adversarial Networks For Launching and Thwarting Adversarial Attacks on Network Intrusion Detection Systems} 
A Generative adversarial network (GAN) - based adversarial attack was proposed by Usama et al. \cite{usama2019generative}. Their method was the first attempt to utilize GAN-based adversarial attacks against a black box Intrusion detection system (IDS) while still preserving the functional behavior of the network traffic. In some cases, their attack dropped the accuracy of the detection model from 84.3 to 43.4 percent.

\subsubsection{Adversarial deep learning for robust detection of binary encoded malware}

Al et al. \cite{al2018adversarial}, developed four adversarial attack methods to generate an adversarial example of a binary malware file that preserves its functionality (rFGSM, dFGSM, BCA, and BGA). They developed a framework for training robust malware detection models by utilizing the saddle-point formulation that consists of the inner maximization and outer maximization problems. The inner maximization approach is used to generate powerful adversarial examples that maximize the loss, and then they inject them in the training time. In some conditions, their attack's evasion rate exceeded 99 percent.

\subsubsection{Investigating Adversarial Attacks against Network Intrusion Detection Systems in SDNs} 
With the increasing deployment
of ML-based NIDSs which leverage the global network visibility offered by SDNs, the threat of vulnerability of the ML
algorithms to adversarial attacks is also considered. Their study considered a use-case example of a SYN Flood DDoS attack, in which they demonstrated the ability to reduce the NIDS detection accuracy from 100\% to
0\% on multiple classifiers using evasion attacks. This was one of the most successful attempts of adversarial attacks against Network Intrusion Detections Systems, proposed by Aiken et al \cite{aiken2019investigating}. Their experimental platform was based on ML based NIDS for Software defined networks called Neptune. In their study, they demonstrated that with the perturbation of a few features, the detection accuracy of a specific SYN flood Distributed Denial of Service (DDoS) attack by Neptune decreases from 100\% to 0\% across a number of classifiers. Furthermore, they proposed an adversarial test suite named Hydra to evaluate the impact of adversarial evasion classifiers against an anomaly-based NIDS - Neptune. Their study considered several classifiers and machine learning algorithms, proving that clustering algorithms were more robust to adversarial samples compared to other ML types. Specifically, KNN proved to be the most robust classifier
against the adversarial attacks performed within their research,
with only one combination of feature perturbations halving the
detection accuracy from 100\% to 50\%. In contrast, Random forest, LR,
and Support vector machines were generally vulnerable to the same perturbations
resulting in similar detection accuracy reductions. The concept of attack generalization was also studied in this publication, using their Neptune NIDS framework as the adversarial target and which was capable of implementing multiple classifiers.

\subsubsection{IoT Network Security from the Perspective of Adversarial Deep Learning}. The effect of adversarial attacks on wireless sensor networks was studied by Sagduyu et. al. \cite{sagduyu2019iot}. The study experimented with adversarial attacks within the context of three types of over-the-air (OTA) wireless attacks, namely within the jamming, spectrum poisoning, and priority violation attack. Their study demonstrated how adversarial attacks can lead to significant loss in throughput, by fooling an IoT transmitter into making a wrong transmit decision in the test phase. This was also an evasion attack against the machine learning model. In their study, they considered an IoT network where an IoT transmitter predicts if a channel status is idle or busy, by using deep learning algorithms. Their study showed that deep learning was effective in performing this task. Then, adversarial machine learning as applied in three contexts - jamming, spectrum poisoning and priority violation attakcs. A defense system based on stackelberg game showed to be an effective mitigation against adversarial machine learning against ioT networks. This defense technique is however considered not  transferable as it was not proven to be generalizable across multiple adversarial attack scenarios.

Several uses of deep learning for anomaly detection in wireless communication systems have been commonly implemented including channel decoding, \cite{liang2018iterative}, wireless resource allocation \cite{sun2017learning} \cite{o2016convolutional} and radio signal (modulation) classification \cite{o2018over}. Uses of Machine Learning in IoT include anomaly detection \cite{canedo2016using}, device identification \cite{meidan2017profiliot} \cite{miettinen2017iot}, and signal authentication \cite{feinman2017detecting}.

\subsubsection{Adversarial Attacks on Deep-Learning Based Radio
Signal Classification} The robustness of deep learning based algorithms for the wireless physical layer was also studied within the context of radio radio signal (modulation) classification tasks. Sadeghi \cite{sadeghi2018adversarial} investigated the use of convolutional neural networks in which they developed both white-box and blackbox adversarial attacks for a DL based modulation classification. In their study, a VT-CNN was used as the classifier. The outcome of their research showed that Significantly less transmit power is required by the attacker in order to
cause misclassification in the case of adversarial machine learning, as compared to the case of conventional jamming (where the attacker transmits only random noise). Hence, adversarial machine learning is an alternative to signal jamming with random noise, with less resource required in terms of transmit power. Their research also created a   a computational efficient algorithm for crafting universal adversarial perturbations (UAP), which can cause a misclasification of the deep learning model irrespective of the input provided to the model. Furthermore, their study revealed an interesting property known as the Shift invariant Property of their attack method, which makes the attack generalizable across various deep learning models, without having any knowledge of the nature of the model, thus implying a black-box attack. Their tests showed that after applying these attacks, the targeted model accuracy could drop from 75 to 0 percent in the cases of a high perturbation-to-noise ratio (ratio of the perturbation power to the noise power).

\subsubsection{Addressing Adversarial Attacks Against Security Systems Based on Machine Learning}
Apruzzese et al. \cite{apruzzese2019addressing} proposed an attack and defense method against several types machine learning algorithms in for network intrusion detection systems. In their study, they evaluated both poisoning and evasive adversarial attacks against three supervised machine learning algorithms. The three algorithms namely Random forest, K-nearest neighbour and Artificial Neural Network (multi-layer perceptron) MLP were used to develop a network intrusion detection system. Their poisoning and evasion attack severity averaged 70.1 and 66.4 percent, respectively. They also demonstrated that adversarial training was effective in improving the robustness of deep learning based network intrusion detection systems.

\subsubsection{Adversarial Deep Learning for Cognitive Radio Security: Jamming Attack and Defense Strategies}
 Shi et al. \cite{shi2018adversarial} proposed an adversarial machine learning approach to launch jamming attacks on wireless communications and introduces a defense strategy. 
The study bases on the premise that in a cognitive radio network,
a typical transmitter workflow includes the task of sensing available channels, identifying spectrum opportunities,
and then transmitting data to the receiver in idle channels. As machine learning techniques have been progressively applied in this context, such as implementing a deep learning classifier for the classification of channels as either idle or busy, attackers seek to compromise the machine learning classifier. Even though the attacker has no knowledge of the deep learning classifier, i.e this is a black box attack. Their experiments showed that their adversarial deep learning attack reduced the transmission success rate from 73.79 to 2.91 percent. The authors also propose a defense technique for the deep learning classifier that works by allowing the transmitter
to deliberately takes wrong actions in predetermined time slots in order to mislead the adversary. 

\subsubsection{Performance Evaluation of Physical Attacks against E2E Autoencoder over Rayleigh Fading Channel} 
Albaseer et. al \cite{albaseer2020performance} investigated the vulnerabilities of autoencoder E2E with Rayleigh channel mode. Their study demonstrated the vulnerability of auntoencoder deep learning models to adversarial samples when used in end-to-end wireless communication systems. Both white-box and black box attacks were launched against and e2e model that was based on a realistic channel model. Their results showed that adversarial attacks had more significant impacts compared to jamming attack.

\subsubsection{Physical adversarial attacks against end-to-end autoencoder communication systems} Sadeghi et al. \cite{sadeghi2019physical} also showed that end to end learning of wireless communication systems are vulnerable to physical adversarial attacks. Similar to the work of  Albaseer et al. \cite{albaseer2020performance}, their study demonstrates that adversarial attacks are more destructive than jamming attacks.

\subsubsection{Targeted Adversarial Examples Against RF Deep Classifiers} Kokalj-Filipovic et al. \cite{kokalj2019targeted} studied the effect of adversarial samples on machine learning based classifiers for radio frequency signals. The goal of their research was to verify if adversarial samples against machine learning based classification in of radio frequency signals was as effects in the physical world (i.e when launched over the air - OTA) as it was in theoretical settings.

\subsubsection{Deep Learning-Based Intrusion Detection With Adversaries}
Wang et al. \cite{wang2018deep} evaluated the vulnerabilities of deep learning-based IDS among state-of-the-art adversarial attack algorithms, including FGSM, JSMA, Deepfool, and CW using NSL-KDD dataset. They recognize feature patterns for the attack algorithms, and they demonstrated that modifying a limited number of features is better for most of the adversaries, such as JSMA attacks. JSMA attacks distinguish adversaries in terms of applicability. They noticed how feature selection to be perturbed by an adversary varies depending on the degree of significance.

\subsubsection{Evaluating Deep Learning-Based Network Intrusion Detection System in Adversarial Environment}
Peng et al. \cite{peng2019evaluating} evaluated the developed scalable ENIDS framework robustness in the adversarial environment against various attacks ( MI-FGSM, L-BFGS, PGD, and SPSA) using NSD-KDD dataset. They compare different well-known models, including SVM, RF, and LR, with the proposed framework under adversarial attacks. They use different metrics to compare the model robustness, including accuracy(ACC), Precision Rate (PR), Recall Rate (RR), F-Sorce (FS), and Success Rate (SR).

\subsubsection{Analyzing adversarial attacks against deep learning for intrusion detection in IoT networks}
Ibitoye et al. \cite{ibitoye2019analyzing} studied the adversarial samples effectiveness against
deep learning-based Intrusion Detection System (IDS) within the context of an IoT network. The authors provide a comprehensive comparison between two different deep learning model, a Self-normalizing Neural Network (SNN) and a  Feed-forward Neural Network (FNN). They utilize and study input features normalization in a deep learning-based IDS in an adversarial environment. It increases the robustness of the deep learning model against various adversarial attacks (FGSM, BIM, and PGD).

\subsubsection{Online anomaly detection under adversarial impact}
Kloft et al. \cite{kloft2010online} studied the effect of a poisoning attack of training data on online centroid anomaly detection (IDS) with a finite sliding
window. They study the poising attack with limited and full control of the training dataset using real HTTP traffic from a web
server of Fraunhofer FIRST institute. This study shows if the attacker has full control of the data, is it easy to attack while when applying additional constraints to have limited control of the training data by assuming that attacker can inject a small fraction of the training dataset, the attack fails. Therefore, adding those constraints adds protection approaches against poising attacks. Their results show that they cannot consider their method secure if the attacker has full control of the dataset. 

\subsubsection{Security evaluation of pattern classifiers under attack}
Biggio et al. \cite{biggio2013security} proposed a framework for empirical security evaluation that can be applied in different three real-life applications, including Intrusion detection system, spam filtering,  Biometric Authentication. They proposed an algorithm to sample training and testing sets.  They evaluate their framework performance under causative adversarial attack using SVM and LR algorithm. For IDS, they used a public data set of a web-server with 205 malicious samples collected in five days in 2006.  Authors recommend the designer of classifiers to follow to use their framework to evaluate the security of the classifier. 

\subsubsection{Evading Machine Learning Botnet Detection Models via Deep Reinforcement Learning}

Wu et al. \cite{wu2019evading} introduced a generic black-box attack against botnet detection machine learning models. The authors of this paper use deep reinforcement learning (DRL) to generate adversarial traffic flows to deceive the detection models. A reinforcement learning agent updates the adversarial samples to change the temporal and spatial features of the traffic flows without altering the original functionality and executability. Their attack's evasion rate ranged from 69.3 to 80.4 percent. 

\subsubsection{Attack-GAN}

Cheng et al. \cite{cheng2021packet} proposed Attack-GAN to generate malicious adversarial raw packets that can mislead current machine learning network intrusion detection systems in the internet of things. Each byte in a packet is represented with word embedding. Feedback from the victim NIDS is needed by this black box attack to update the parameters of the generator. The attack success rate depends on multiple factors like the machine model and the modes of byte embedding, but it reached 98.42 percent in the best case.

\subsubsection{Fooling intrusion detection systems using adversarially autoencoder}

Chen et al. \cite{chen2021fooling} introduced AIDAE (Anti-Intrusion Detection AutoEncoder) framework against IDSs. AIDAE can produce features matching normal feature distribution, it also keeps the correlation between the generated continuous and discrete features. They used Evasion Increase Rate (EIR) to evaluate their attack. The EIR reflects the evasion power by comparing the adversarial detection rate with the original, i.e. 1-(adversarial detection rate/original detection rate). EIR was higher than 0.9 in all their experiments. 

\subsubsection{TANTRA}

Sharon et al. \cite{sharon2021tantra} presented TANTRA (Timing-Based Adversarial Network Traffic Reshaping) which deceives NIDSs by reshaping attack network traffic using the timestamp attribute. Based on the authors' evaluation, TANTRA had an extremely high success rate (99.99 percent). However, when TANTRA was tested after training the NIDSs with both benign and reshaped traffic, its success rate decreased.

\begin{figure*}
	\includegraphics[width=\linewidth,keepaspectratio=true]{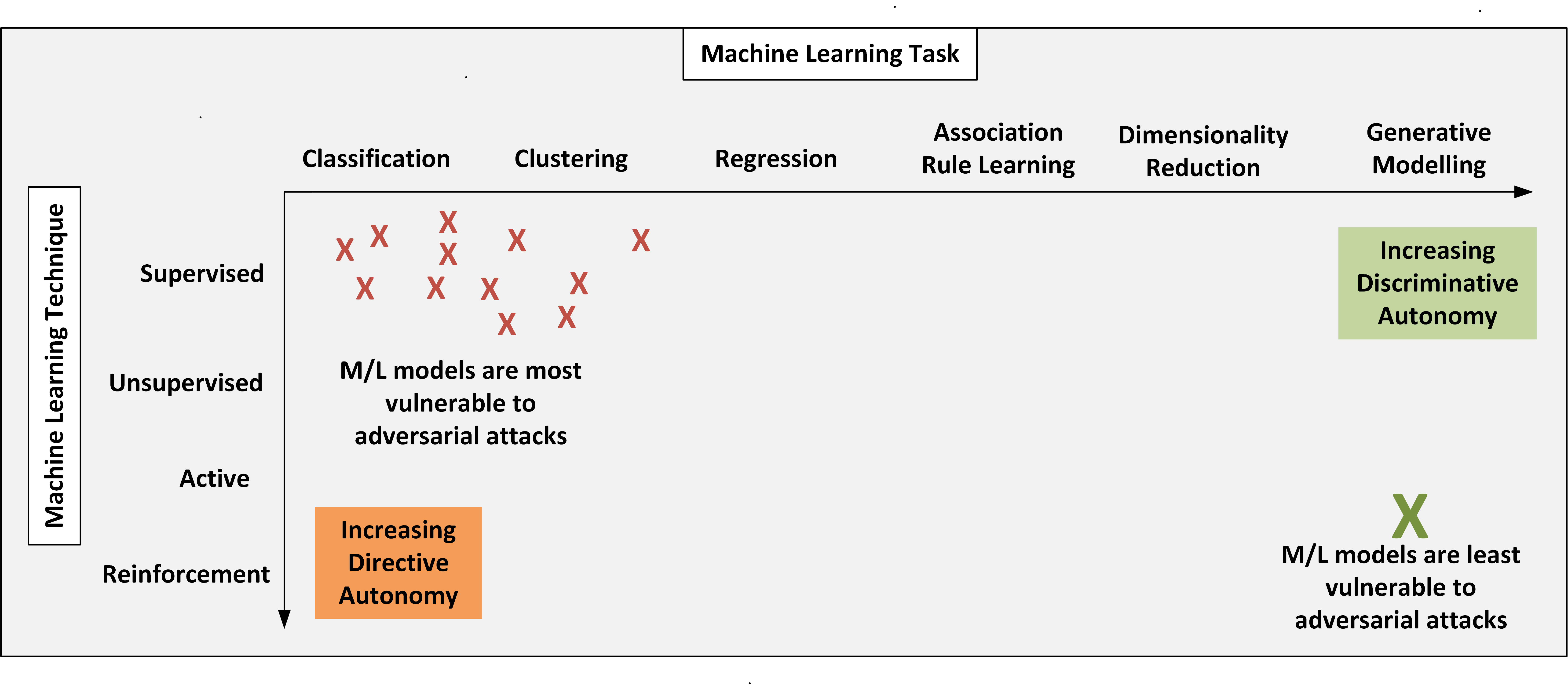}
	\caption{Adversarial Risk Grid Map}
	\label{fig:adv_risk}
	\centering
\end{figure*}

\section{Evaluating Adversarial Risk}
In discussing adversarial risk, we introduce the concept of discriminative and directive autonomy of machine learning models. The two-fold goal of an adversarial risk grid mapping is to evaluate the likelihood of success of an adversarial attack against a machine learning model, and the consequence of that attack if successful. Adversarial risk often seek to measure the performance of a machine learning model based on worst case inputs \cite{uesato2018adversarial}. We present in this paper, an adversarial risk grid map shown in Figure \ref{fig:adv_risk} based on the level of autonomy of the machine learning model with respect to the learning technique and task. The concept of discriminative autonomy and directive autonomy of the machine learning models represents a novel approach for evaluating the relative adversarial risk of a machine learning model.

\subsection{Security by Obscurity in Adversarial Risk}
The notion of security by obscurity in adversarial context, in which defenses are proposed based on obscurity to an adversary does not truly reflect the nature of adversarial risk in machine learning-based network security applications. The prevalence of black box adversarial attacks which fool classifiers without having direct access to the model further demonstrate the weakness in the obscurity approach to adversarial risk.

As adversarial attacks continue to emerge into real world production systems, the ability to computationally evaluate and even optimize adversarial risk becomes invaluable. While both adversarial risk and obscurity have been impossible to compute directly  \cite{uesato2018adversarial}, frameworks for adversarial risk based on the concept of obscurity have been proposed \cite{liao2018defense}.

\subsection{Adversarial Risk Grid Map}
A modified notion of adversarial risk was proposed in \cite{suggala2019revisiting} which suggested that certain classifiers inherently have low adversarial risk. Other works \cite{fawzi2018adversarial} \cite{tsipras2018there} have suggested a trade-off between standard risks and adversarial risk. This indicates that with increase in standard accuracy of the classifier, the adversarial risk of the classifier increases. Based on our review, a grid map based on the autonomy of the machine learning model is proposed. We term this as model autonomy adversarial risk approach since it is based on the directive and discriminative autonomy of the machine learning models. The map is shown in Figure \ref{fig:adv_risk}.

\begin{itemize}
\item \emph{Discriminative Autonomy:}
The discriminative autonomy is directly related to the type of task being performed by the machine learning model. Machine learning tasks such as classification are highly dependent on the input data. As such, they have lower discriminative or conditional autonomy compared to tasks such as generative modeling which depend less on the input data when predicting an outcome.

\item \emph{Directive autonomy:}
The directive autonomy of a machine learning model is a function of the machine learning technique. In supervised machine learning, there is less directive autonomy since the model needs to be first learned with some form of labeled data. Machine learning techniques such as reinforcement learning depend less on a model being learned with any form of training data and posses much higher directive autonomy.
    
\end{itemize}

\subsection{Cross Model vs Cross Dataset Attack}
In discussing adversarial risk, the notion of transferability becomes pertinent. Transferability refers to the fact in which an adversarial example which is crafted for a specific deep learning model, is found to be effective in causing a misclassification in a different model. This is known as cross-model adversarial samples. in a similar situation, when the adversarial sample that was generated by altering a particular dataset. If that sample is used to attack a deep learning system that was trained using a different dataset, that is called a Cross-dataset adversarial sample.

\section{Defending Against Adversarial Attacks}

Numerous researchers have aimed to review and classify defenses against adversarial attacks. Barreno et al. \cite{barreno2006can} first proposed three broad approaches for defending machine learning algorithms against adversarial attacks. Regularization, Randomization, and Information hiding. Yuan et al. \cite{yuan2017adversarial} classified the defenses into two broad strategies. Proactive strategies and reactive strategies. Rosenberg et al. \cite{rosenberg2021adversarial} organized the defenses based on the cyber security sub-domains (malware detection, spam detection, biometric systems, etc.), in our work, we classify the defenses based on generalized ML approaches.

Since adversarial examples represent a worst-case scenario of a distribution shift, the task of generating an adversarial sample is a non-convex optimization problem that can only be approximately solved. Adversarial attack methods are mostly optimization algorithms in search for a lower boundary perturbation that corresponds to an adversarial sample \cite{brendel2017decision} . These optimization algorithms often result in high frequency outputs \cite{guo2018low}. This however makes the defense methods against this adversarial samples vulnerable to adversarial samples that are generated within a low-frequency subspace. 

In this section, we provide the most common defense methods in use today and classify them based on the strategy and approach. The reviewed defense methods are shown in Figure \ref{fig:defense_strategy} below.

\subsubsection{Gradient Masking} 
Since most method of adversarial attacks are based on the using of gradient, the gradient masking method modifies a machine learning model in an attempt to obscure its gradient from an attacker. Nayebi et al \cite{nayebi2017biologically} demonstrated the effect of gradient masking by saturating the sigmoid network which results in a vanishing gradient effect in gradient-based attacks.  Authors force the neural networks to works in nonlinear saturating system. By using Jacobian regularization for each network layer including the output layer, the model becomes non sensitive of perturbations that are generated using fast gradient sign method (FGSM) and iterative adversarial attacks\cite{nayebi2017biologically}.
However, \cite{yanagita2018gradient} indicate that gradient masking react as over-fitting in their experiments.

\begin{figure*}

\centering
\begin{forest}
for tree={
    draw, 
     text centered 
    }
[ Adversarial Defense Methods
[Gradient\\Masking\\\cite{nayebi2017biologically} ]
[Defensive\\Distillation\\\cite{hinton2015distilling}]
[Adversarial\\Training\\\cite{madry2017towards}]
[Gradient\\Regularization\\\cite{ross2018improving}]
[Detecting\\Adversarial Samples\\\cite{feinman2017detecting}]
[Feature\\Reduction\\\cite{grosse2016adversarial}]
[Input\\Randomization\\\cite{xie2017mitigating}]
[Ensemble\\Defenses\\\cite{song2017pixeldefend}]
]
\end{forest}

\caption{Adversarial Defense Methods}
\label{fig:defense_strategy}
\end{figure*}

\subsubsection{Defensive Distillation}
Distillation technique was originally proposed by Hinton et al. \cite{hinton2015distilling} for transferring knowledge from large neural networks to smaller ones. To implement the distillation approach, Hinton et al. authors built 10 DNN models with same architecture and training method and use soft targets to avoid overfitting that occur when using hard targets.  They proved in their experiments that ensemble model is able to transfer knowledge to the distilled model better than individual models. However,  ensemble requires large computation models that have large networks and large datasets. Therefore, they use learning specialist models that each use a subset of dataset classes to reduce the amount of computation \cite{hinton2015distilling}.
Also, it was adapted by Papernot et al. \cite{papernot2016distillation} to defend against adversarial crafting by using the output of the original neural network to train a smaller network rather than using the distillation as originally proposed by Hinton. Defensive distillation was initially tested against adversarial attacks in computer vision, but further research is required to determine its effectiveness in other applications such as malware detection.

\subsubsection{Adversarial Training}
Adversarial training \cite{madry2017towards} is a method that aims to increase the robustness of a machine learning model to adversarial samples by minimizing the loss \(L\) on data/label pairs \(\{{X_i},{y_i}\}\) while maximizing the corresponding loss function. Szegedy et al. \cite{szegedy2013intriguing} originally proposed a three-step method known as adversarial training for defending against adversarial attacks. 1, Train the classifier on the original dataset 2, Generate adversarial samples 3, Iterate additional training epochs using the adversarial samples. Generally, adversarial training is based on min-max formulation that solves two problems: attacks as an inner maximization problem and defenses as an outer minimization problem to achieve optimization \cite{madry2017towards}. The inner maximization intents to generate adversarial samples version that results to maximize the model loss. Where the outer minimization intents to minimize the loss by finding model parameters that build a more robust model with less adversarial loss \cite{madry2017towards}. Numerous researchers tested and evaluated the effect of adversarial training in the network security domain \cite{abou2019investigating,abou2020evaluation}. They concluded that it improves the classification performance of the machine learning model and makes it more resilient to adversarial crafting.

However, adversarial training has certain limitations particularly in the context of adversarial machine learning in network security. First, the adversary may implement a different attack method other than the one which was used in training the
network. Secondly, the adversary may design adversarial perturbations for a deep learning model that already has been trained with
adversarial training, and craft new adversarial perturbations which would make the previous adversarial training ineffective.
It has also been shown that adversarial training can reduce the performance of the deep learning models on clean inputs as discussed in \cite{sadeghi2019physical}.

\subsubsection{Gradient Regularization}
Gradient regularization is a technique that penalizes large changes in the output of some neural network layer, to adjust machine learning models, minimize the loss function, increase model robustness and prevent overfitting or underfitting. Many researchers tested this approach as a defense against adversarial attacks, like Ros et al. \cite{ross2018improving} who found that training DNNs with gradient regularization improves the robustness to adversarial perturbations as much or more than adversarial training. They have also found that combining both approaches (gradient regularization and adversarial training) achieves greater robustness. The main drawback of Gradient regularization is that it doubles the training time per batch.

\subsubsection{Detecting Adversarial Samples}
Several approaches are used to detect the presence of adversarial samples in the training phase of a machine-learning model. One of such approaches proposed by \cite{feinman2017detecting} works on the premise that adversarial samples have a higher uncertainty than clean data and uses a Bayesian neural network that is in dropout layers of neural networks to estimate the extent of uncertainty in the input data to detect the adversarial samples. Other approaches include the use of probability divergence proposed by \cite{meng2017magnet} as well as the use of an auxiliary network  of the original network introduced by Metzen et al. in \cite{metzen2017detecting}. Ren et al. \cite{ren2022towards} also proposed adversarial attack detection and adversarial sample recognition methods by using the causal inference technique to establish a causal model to describe the generation and performance of adversarial samples that attack DNNs.

\subsubsection{Feature Reduction}
Other potential defenses for adversarial attacks have been proposed. Simple feature reduction was evaluated by Grosse et al. \cite{grosse2016adversarial} but was found inadequate in defending against adversarial attacks. 

\subsubsection{Input Randomization}
Some researchers tried randomization operations on the input of the model as a defense against adversarial attacks on machine learning, for example, Xie et al. \cite{xie2017mitigating} tried random resizing and adding random padding on inputs. Experiments demonstrated that their proposed method is effective. Zhang et al. \cite{zhang2019defending} also tried a similar method by injecting random Gaussian noise. These approaches have multiple advantages such as simplicity, low computational complexity, and eliminating the need for additional training. The main disadvantage is using this defense technique in the network security domain could change the functionality of the inputs (Executables, Packets, etc.). In our opinion, this method needs to be evaluated in the network security domain.

\subsubsection{Ensemble Defenses}
Similar to the idea of ensemble learning which combines one or more machine learning techniques, researchers have also proposed the use of multiple defense strategies as a defense technique against adversarial samples. PixelDefend was proposed by \cite{song2017pixeldefend} to combine adversarial detecting techniques with one or more other methods for creating a more robust defense against adversarial attacks.





\section{Discussion and Lessons Learnt}
This section discusses several key lessons learnt through our survey on adversarial attacks against ML in network security.
\subsection{Increased Adversarial Risk}
We observed an increased risk of adversarial vulnerability of machine learning models in network security with reduced discriminative autonomy and directive autonomy. Similarly, we observed a reduced risk of adversarial vulnerability with increased discriminative autonomy and directive autonomy. As illustrated in the adversarial risk grid map shown in Figure \ref{fig:adv_risk}, the discriminative autonomy directly relates to the machine learning tasks while the directive autonomy relates to the machine learning technique. The reason for the adversarial sensitivity  of the machine learning models to the discriminative and directive autonomy based risk grid map is still an area of open research.

Previous approaches on making machine learning in network security more secure have advocated the development of machine learning models that are resilient to adversarial attacks. In this survey, we introduced the concept of an element of reduced risk of adversarial attacks based on an adversarial risk grid map. Our findings suggest that the adversarial risk grid map provides a promising future for the security of artificial intelligence and machine learning in network security. Machine learning based network security applications that are more resilient to adversarial attacks can be designed by leveraging on the adversarial risk grid map. We observed that the misclassification achieved by an adversarial attack is dependent significantly on the design of the adversarial attack algorithm with the context of each specific attack . White-box, Evasion attacks against endpoint protection systems (malware detection) are the most common attacks. While there is limited research in adversarial attacks against process behavior and user behavior analysis, use cases of machine learning in network security, endpoint protection, network protection and application security have been well researched.

\subsection{Transferability with regards to machine learning technique}Transferability of adversarial samples \cite{papernot2016transferability} \cite{tramer2017space} has been shown to be more effective with targeted adversarial samples \cite{liu2016delving}. This implies that non-targeted adversarial samples (reliability attacks) which are solely aimed at causing a misclassification, are more likely to transfer from one model to the other. In furtherance to this phenomenon, we observe that adversarial attacks in network security are less likely to transfer from one machine learning technique to another. Transferability of adversarial defences in network security in also impacted by to the heterogenous nature of the perturbed features. While this is has a positive side with regards to preventing transferable defenses, it also makes it more difficult in real world situations. From our observation, adversarial attacks in problem space are more difficult to generate, more difficult to defend against and less chances of being transferable. 

In our research, we observed that a significant amount of features are perturbed in the process of generating the adversarial sample. This is a sub optimal approach. There is currently no publication which has explored the challenge of finding a way to identify the ideal features that need to be perturbed for creating adversarial samples. In the field of computer vision, Guo et al. \cite{guo2018low} restricted the search for adversarial samples to the low frequency domain, thereby reducing query complexity.

We reviewed defenses against adversarial attacks on machine learning applications in network security. We note that there are two major limitations in the existing research on adversarial defenses. Firstly, most defenses are designed to protect against attacks on machine learning applications in computer vision. Secondly, the defenses studied are usually designed for a specific attack or a part of the attack. A generalized defense model against adversarial attacks is at best still theoretical as research on generalized defense models is in early stages \cite{schmidt2018adversarially}. Furthermore, our findings indicate that  defenses against adversarial attacks are specific to a particular type of attack and are not necessarily transferable. Recent research \cite{papernot2016transferability} have studied the transferability in malware machine learning models in machine learning applications such as malware detection.

\subsection{Malware Detection Approaches} In the majority of cases, Android malware detection is posed as a binary classification problem in which a classifier is used to determine whether an app is malicious or not.
Malware detection take three general approaches which are dynamic, static, or hybrid. Significant overhead is usually required in order to extract dynamic features because it requires monitoring the behavior of apps at run time. Several of the studies we examined have focused on instances in which static features were extracted, including required permissions, actions, and application programming interface (API) calls. In our literature review, we did not come across any work in which adversarial attacks were successfully carried out against machine learning based malware detection systems in which dynamic features were extracted.

\subsection{Quantitative evaluation of adversarial attacks} In network security, majority of the adversarial attacks reported target the integrity aspect of the CIA triad, with the intent of causing a misclassification. A quantitative analysis of the attacks' efficiency for the four reviewed categories (malware detection, phishing detection, spam detection, and network anomaly detection) was observed.   
After calculating the average attack success rate per class, we have found that the most significant adversarial effect was in the malware detection and the network anomaly detection domains, in which the adversarial attacks' success rates averaged more than 90 percent. It is worth mentioning that we think that a number of these attacks are theoretical and need more investigation to deploy them in practical settings, thus the quantitative effect of some of the reviewed attacks could be exaggerated. However, we find these results as a good indication of the malicious potential of adversarial attacks on network security domains.

The challenge of quantifying the efficiency of adversarial example generation, is an emerging field and several approaches have been proposed in recent literature. In \cite{li2019adversarial} a new performance metric was proposed, called effective generation rate (EGR) which is the ratio between n~ and n, i.e., n~/n. Where n represents the number of adversarial examples generated by an attacker and n~ denotes the number of adversarial examples that successfully evades both malware and adversarial example detection.

\subsection{Difference between adversarial attacks in network security and computer vision}
\begin{enumerate}
    \item In image recognition, the primary feature used in adversarial perturbation is the pixels of the image. However, in network security, there is a great variation in the types of features which may be used, and as such, the perturbation scope for adversarial attacks becomes largely increased.
    
    \item Adversarial attacks in network security differs from computer vision since data objects are considered rather than images. As a result, the perturbed features are more diverse and heterogeneous. The consequence of this is that it becomes more difficult to defend against adversarial attacks in network security due to the heterogeneity , hence, transferability and universal defenses against . It should be noted that significant strides have been made in computer vision, with regards to developing universal defences but this is still an infant research area in network security. Also, the feature varies greatly based on the network security application. In most cases, the features used in the machine learning classification are also the features that are perturbed in generating the adversarial samples. 

\end{enumerate}

\section{Conclusion and future Work}

We present a first of its kind survey on adversarial attacks on machine learning in network security. The previous survey \cite{akhtar2018threat} that we reviewed had only discussed adversarial attacks against deep learning in computer vision. We introduced a new classification for adversarial attacks based on applications of machine learning in network security and developed a matrix to correlate the various types of adversarial attacks with a taxonomy-based classification to determine their effectiveness in causing a misclassification. We also presented a novel idea of the concept of an adversarial risk grid map for machine learning in network security.

 In our review on defenses against adversarial attacks, although there were numerous proposed defenses against specific adversarial attacks, research on generalized defenses against adversarial attacks is still not well established \cite{schmidt2018adversarially}. In our future work, we would study generalized defenses against adversarial attacks to understand if a generalized approach towards adversarial defenses will be effectively attainable. In addition, we would examine the interpretability of the adversarial risk to further understand why the reduced adversarial vulnerability occurs, and its implications for other applications of machine learning such as computer vision and natural language processing.

\textbf{Future Work}
Based on our research, adversarial attack has mostly been carried out on data at rest, with very few successful attempts of adversarial attacks on data in transit, or streaming data such as \cite{xie2020real} \cite{gong2019real}. However, in network security domains such as in the field of intrusion detection, realistic adversarial attacks will be carried out on data in transit. Hence, more research is needed in this area to understand the potential risks of adversarial attacks against data in transit and the possible defense techniques. 

Adversarial attack against federated learning \cite{bagdasaryan2020backdoor}  is still an open area of research. Federated learning \cite{yang2019federated}, which is quite different from distributed computation, involves the situation in which each client performs the machine learning computation without sending the data to the cloud. As such, the cloud provider does not have a complete view of the machine learning model with significant gains for privacy and confidentiality. 

Adversarial attacks were demonstrated to affect only classifier and clustering tasks in network security. From the reviewed literature of over fifty attacks against machine learning in network security, there has been no attempt to implement adversarial attacks against any other task in network security except classification and clustering tasks. This is consistent with our adversarial risk grid map illustrated in Figure \ref{fig:adv_risk} in which we posit that adversarial risk increases based on the type of network security task which is being performed. Our study notes that there are diverse adversaries in network security compared to computer vision. as such, there is even more relevant arms race situation in network security than in computer vision

Several authors have shown that deep learning can be performed on data that is encrypted\cite{aceto2018mobile} \cite{hesamifard2017cryptodl} \cite{lotfollahi2020deep}. But in our study, we observe that encrypted data has not been adversarial defeated. Even though, most data in network security is encrypted, adversarial attacks or the ability to generate adversarial samples against encrypted data is an area of open research. As such, it is a promising idea, subject to future research, to stipulate that performing encryption before applying machine learning to the data, is a trusted and proven defense against adversarial machine learning in network security.

The use of deep learning as a technique for encryption is quite restrictive \cite{klein2005synchronization}. This is mostly due to the computational costs of deep learning. Research is also required to understand the effects of adversarial attacks against deep learning for encryption.

\section*{Acknowledgement}
This work was supported by the Natural Sciences and Engineering Research Council of Canada (NSERC) through the NSERC Discovery Grant program.

\bibliographystyle{ieeetr}
\bibliography{main}

\end{document}